\documentclass{article}
\begin{document}

\def\sa{\section}
\def\sb{\subsection}
\def\sc{\subsubsection}

\def\ind{\ \ \ \ }

\def\be{\begin{equation}}
\def\ee{\end{equation}}

\def\bea{\begin{eqnarray}}
\def\eea{\end{eqnarray}}

\def\ba{\begin{array}}
\def\ea{\end{array}}

\def\nn{\nonumber}

\def\ben{\begin{enumerate}}
\def\een{\end{enumerate}}

\def\fn{\footnote}

\def\rd{\partial}
\def\rot{\nabla\times}

\def\r{\right}
\def\l{\left}

\def\gt{\rightarrow}
\def\cf{\leftarrow}
\def\bw{\leftrightarrow}

\def\ra{\rangle}
\def\la{\langle} 
\def\bla{\big\langle}
\def\bra{\big\rangle}
\def\Bla{\Big\langle}
\def\Bra{\Big\rangle}

\def\ddt{{d\over dt}}

\def\rdt{{\rd\over\rd t}}
\def\rdx{{\rd\over\rd x}}

\def\bb{\begin{thebibliography}{99}}
\def\eb{\end{thebibliography}}
\def\bit{\bibitem}

\def\bc{\begin{center}}
\def\ec{\end{center}}

\title{Noncommutative geometry and nonabelian Berry phase 
in the wave-packet dynamics of Bloch electrons}

\author{Ryuichi Shindou
\fn{Department of Physics, University of Tokyo,
Hongo 7-3-1, Bunkyo, Tokyo 113-0033, Japan.},
Ken-Ichiro Imura
\fn{Condensed Matter Theory Laboratory,
RIKEN (Wako), Hirosawa 2-1, Wako, 351-0198, Japan.}
\fn{corresponding author (e-mail: imura@riken.jp).}}

\date{\today}

\maketitle

\begin{abstract}
Motivated by a recent proposal on the possibility of observing a 
monopole in the band structure, and by an increasing interest 
on the role of Berry phase in spintronics, we studied the adiabatic 
motion of a wave packet of Bloch functions, under a perturbation 
varying slowly and incommensurately to the lattice structure. 
We show, using only the fundamental principles of quantum mechanics, 
that the effective wave-packet dynamics is conveniently described 
by a set of equations of motion (EOM) for a semiclassical particle 
coupled to a {\it nonabelian} gauge field associated with a geometric 
Berry phase.

Our EOM can be viewed as a generalization of the standard Ehrenfest's 
theorem, and their derivation was asymptotically exact in the 
framework of linear response theory. 
Our analysis is entirely based on the concept of {\it local} 
Bloch bands, a good starting point for describing the adiabatic 
motion of a wave packet.
One of the advantages of our approach is that the various types
of gauge fields were classified into two categories by their 
different physical origin: 
(i) projection onto specific bands, 
(ii) time-dependent {\it local} Bloch basis. 
Using those gauge fields, we write our EOM in a covariant
form, whereas the gauge-invariant field strength
stems from the {\it noncommutativity} of covariant derivatives 
along different axes of the reciprocal parameter space.
On the other hand, the degeneracy of Bloch bands makes 
the gauge fields {\it nonabelian}.

For the purpose of applying our wave-packet dynamics to the analyses 
on transport phenomena in the context of Berry phase engineering, 
we focused on the Hall-type and polarization currents. 
Our formulation turned out to be useful for investigating and 
classifying various types of topological current on the same footing. 
We highlighted their symmetries, in particular, their behavior under 
time reversal ($T$) and space inversion ($I$). 
The result of these analyses was summarized as a set of cancellation 
rules, schematically shown in Tables 1 and 2. 
We also introduced the concept of {\it parity} polarization current, 
which may embody the physics of orbital current. 
Together with charge/spin Hall/polarization 
currents, this type of orbital current is expected to be a potential 
probe for detecting and controling Berry phase.
\end{abstract}

\section{Introduction}
\ind
The search for a quantized magnetic monopole has a long history.
\cite{dirac,pol}
Recently a group of condensed-matter physicists 
\cite{masa,fang} 
embodied the idea of detecting a monopole in the band structure.
\cite{vol} 
In crystal momentum space, monopoles appear 
as a source or a sink of the reciprocal magnetic field 
\cite{CN,SN}
associated with the geometric phase of Bloch electrons.
The geometric phase of a Bloch electron, i.e., 
its Berry phase, has also attracted much attention on the 
technological side, in particular, in the context of {\it spintronics}.
A spin Hall effect has been of much theoretical concern,
\cite{hirsch,zhang,MNZ,sinova,culcer}
since it may provide a possible efficient way to induce spin 
current in a semiconductor sample on which spintronic devices
\cite{DD} will be constructed.

The subject studied in this paper stands at the interface
between the forefront of the search for a monopole 
and the latest technology of spintronics.
We study the wave-packet dynamics of a Bloch electron under
pertubations slowly varying in space and in time.
We derive and analyze a set of equations of motion (EOM)
which describes the center-of-masss motion of such a wave packet 
together with its internal motion associated with
its (pseudo) spin.
A reciprocal gauge field of geometric origin 
(Berry connection) appears naturally in such EOM.
\cite{SN}
Then we combine our formalism with the Boltzmann transport 
theory to describe such phenomena as spin and orbital transport.
Its relevance to quantum charge/spin pumping 
\cite{TN,RS2}
will be also briefly discussed.

Before going further into the detailed descritption of 
our project, let us briefly remind you what the Berry phase is, 
and how it has become to be widely recognized in the community.
In his landmark paper \cite{berry},
Berry introduced it as a quantal phase acquired by a
wave function whose Hamiltonian is subject to an adiabatic
perturbation.
The Berry connection, i.e., a gauge field appears
as a phase of the overlap of two wave functions infinitesimally 
separated in the adiabatic parameter space.
Before being formulated in such a systematic manner,
the Berry phase, however, had already been recognized and 
discussed, for somewhat restricted cases though,
in several independent contexts.
The molecular Aharonov-Bohm effect discussed in Ref. \cite{mead}
is nothing but a manifestation of Berry phase.
Its relevance to band structure had also been recognized
in limited situations, such as anomalous Hall effect (AHE), 
\cite{KL,blout} as well as in the study of
quantized Hall conductance. \cite{TKNN}
The role of Berry phase in ferroelectrics has also been of 
much theoretical interest. \cite{resta}
Recently, the Berry phase in AHE has attracted a renewed
attention, revealing its rich topological structures. 
\cite{masa,fang,CN,SN,OMN,RS1,JNM,yao}
The Berry phase has also been generalized to a non-Abelian 
case. \cite{WZ}

The equations of motion (EOM) for a wave packet of Bloch 
functions 
\fn{A Bloch function is an eigenstate of a periodic
Hamiltonian such as Eq. (\ref{H0}), whose energy spectrum 
forms a band structure $\epsilon_n^{(0)}(\bar{\bf k})$
defined as in Eq. (\ref{en0}).}
is instrumental in all the analyses done in the paper.
In order to illustrate our program, we begin with some
details of the description of such EOM.
A wave packet of Bloch functions is localized in the phase 
space around $(\bar{\bf k},\bar{\bf x})$
(where $\bar{\bf k}$ is a crystal momentum characterizing
the Bloch function).
The wave packet is also composed of a specific Bloch band $n$,
whose energy dispersion relation is given by
$\epsilon_n^{(0)}(\bar{\bf k})$.
The center of mass coordinates $(\bar{\bf k},\bar{\bf x})$
obey a set of classical EOM, as the Ehrenfest's theorem says. 
In the presence of electromagnetic field $(\bf{E,B})$,
its motion is subject to an electric and Lorentz forces,
\bea
{d\bar{\bf k}\over dt}&=&
-e\bigg({\bf E}({\bf x})+{d\bar{\bf x}\over dt}\times{\bf B}
({\bf x})\bigg),
\label{C1} \\
{d\bar{\bf x}\over dt}&=&
{\rd \epsilon_n^{(0)}(\bar{\bf k})\over\rd \bar{\bf k}}.
\label{C2}
\eea
These EOM, together with the Boltzmann transport theory, 
describe the electro-magnetic response of the system. 
To see this point, let us express the charge current 
in terms of the momentum distribution function 
$f(\bar{\bf k})$ as
\be
{\bf J}_{\cal C}=-e \int{d\bar{\bf k}\over (2\pi )^D} 
f(\bar{\bf k}) {d\bar{\bf x}\over dt},
\label{J}
\ee
where $D$ is the dimension of coordinate space.
The net current vanishes in the thermal equilibrium.
A finite net current appears when either 
\ben
\item
$f(\bar{\bf k})$ is deviated from its equilibrium value, or
\item 
$d\bar{\bf x}/dt$ acquires an anomalous term,
i.e., an anomalous velocity.
\een
Case 1 corresponds obviously to the usual ohmic
transport, in which
the current is induced by a small deformation of a Fermi
sphere from its thermally equilibrated distribution.
In this case the current is, therefore, carried only by the electrons
in the vicinity of the Fermi surface.

The Berry phase contribution to Eq. (\ref{J}) corresponds to
Case 2, and involves, in contrast to Case 1, all the electrons 
below the Fermi surface.
This type of geometric current might be also 
{\it dissipationless}. \cite{MNZ,inoue,loss}
When Berry connection is taken into account,
the classical EOM, in particular, Eq. (\ref{C2}) is subject to
a modification.
In terms of a {\it reciprocal magnetic field} ${\cal B}$,
the EOM for $\bar{\bf x}$ now reads,\cite{SN}
\be
{d\bar{\bf x}\over dt}=
{\rd \epsilon_{\rm eff}(\bar{\bf k},\bar{\bf x},t)
\over \rd \bar{\bf k}}
+{d\bar{\bf k}\over dt}\times{\cal B}(\bar{\bf k}),
\label{recipro}
\ee
where $\epsilon_{\rm eff}(\bar{\bf k},\bar{\bf x},t)$ is an effective energy,
which will be defined in more precise terms in Eq. (\ref{eff}).
The nature of reciprocal magnetic field will be clarified in Sec. 2.
One can observe in Eq. (\ref{recipro}) that
${\cal B}(\bar{\bf k})$ acts quite similarly to the Lorentz force in the
real space.
${\cal B}(\bar{\bf k})$ encodes information on the topological
nature of band structure, in particular, that of band crossings.
\cite{SN}
Indeed, a degeneracy point corresponds to a momopole
of ${\cal B}(\bar{\bf k})$, 
\cite{vol}
which has played a crucial role in the
understanding of anomalous Hall effect (AHE)
\cite{CN,OMN,RS1,JNM}.

In this paper we study the wave-packet dynamics of Bloch electrons
subject to a perturbation $\beta({\bf x}, t)$
varying slowly in space and time.
Even though our treatment of $\beta({\bf x}, t)$ is in completely
general terms, we can give some concrete examples of 
$\beta({\bf x}, t)$ as in Ref. \cite{SN},
\be
H({\bf p,x} ; \beta({\bf x},t))=
H_0\big({\bf p}+{\bf \beta_1}({\bf x}, t), {\bf x}+{\bf \beta_2}({\bf x}, t)\big)
+\beta_3({\bf x}, t).
\label{beta}
\ee
$H({\bf p,x}; \beta=0)$ is an unperturbed Hamiltonian.
The first two categories, ${\bf \beta_1}({\bf x}, t)$ and ${\bf
\beta_2}({\bf x}, t)$,
are in a vectorial form, whereas $\beta_3({\bf x}, t)$ is a scalar.
In the case of electro-magnetic perturbations, 
${\bf \beta_2}({\bf x},t)=0$.
In the following, we restrict our analyses, for the sake of simplicity, 
to the case of
\be
{\bf \beta_2}({\bf x},t)=0.
\ee
A finite ${\bf \beta_2}({\bf x},t)$ could be relevant, e.g., 
for the study of deformational perturbations in a crystal.
\cite{SN}

Following the quantum mechanical motion of a wave packet
localized around $(\bar{\bf k},\bar{\bf x})$,
we study its EOM focusing on the topological
nature of band structure, and interpret them in terms of
the reciprocal vector potential ${\cal A}_q$
defined in the $(2D+1)$-dimensional parameter space
$\{q\}=(\bar{\bf k},\bar{\bf x},t)$.
This set of parameters $\{q\}$ plays in our case
the role of adiabatic parameters in the original
formulation of Berry phase. \cite{berry}
Our approach is entirely based on the fundamental relations of
Schr\"odinger qauntum mechanics, and makes no referenece to
(i) Time-dependent variational principle \cite{SN}, or
(ii) Path-integral method using Wannier basis \cite{KT}.
Although our approach is conceptually much simpler than those
mentioned above, this type of analysis can be found,
to our knowledge, only in the classical literature.
\cite{zak,blout}
We have in mind a linear response theory with the help of
Boltzmann equation. We, therefore, restricted our analysis to the first
order of external perturbation $\beta({\bf x}, t)$.
We emphasize here that all our analyses are
{\it asymptotically exact} in the framework of linear response theory.

This paper is organised as follows :
In Sec. 2, we first discuss the nature of non-Abelian
gauge field, appearing in our EOM, which will be derived later
in Sec. 4. In Sec. 3, we state and formulate unambiguously
our problem, as well as listing all the assumptions we will
make. The EOM is derived in Sec. 4, whose possible application
to Berry phase engineering is discussed in Sec. 5,
before coming to the conclusions in Sec. 6.
Some technical details are left for Appendices.

\sa{Origin of the gauge field}

\ind
The nature of a reciprocal magnetic field ${\cal B}(\bar{\bf k})$ 
appeared in Eq. (\ref{recipro}) lies, 
as will be further discussed in Sec. 4, 
in the noncommutativity of the center of mass coordinates,
$(\bar{\bf k},\bar{\bf x})$.
\cite{MNZ}
In more mathematical terms, ${\cal B}(\bar{\bf k})$
is a curvature associated with a geometric Berry connection,
i.e., a gauge field.
The relation between such noncommutative coordinates as seen in
Eqs. (\ref{xx},\ref{dd}) and the MM 
in momentum space has been of much theoretical interest.
\cite{masa,fang,MNZ,Metz}.
From a more general point of view,
physics in noncommutative space-time coordinates
has been of great theoretical interest,
rather in high-energy physics community,
in particular, in the context of string and
$M$ theories.
\cite{seiberg,connes}
In the following we consider, instead, the physical origins 
from which our gauge fields stem, and the mechanism 
how they are generated, focusing on the case of Bloch
electrons under slowly varing perturbation $\beta({\bf x},t)$.

\sb{Non-Abelian gauge field, or Berry phase,
encoding information on the band structure}

\ind
Let us consider the motion of a wave packet
composed of a limited number, say, $N$ of degenerate bands
over the whole Brillouin zone.
When neither the time reversal symmetry nor
the spatial inversion symmetry is broken, there always appears
a two-fold degeneracy at every ${\bf k}$-point (Kramers doublet).
If there is no further degeneracy in the system,
then $N=2$ in our language.
In the following chapters, we will derive,
using only the most fundamental relations of
Schr\"odinger quantum mechanics,
effective equations of motion (EOM) for this wave packet.
These EOM are most conveniently interpreted in terms of
non-Abelian gauge fields in the reciprocal
space.
When we derived these effective equations of motion, we restricted
our available Hilbert space to these degenerated bands.
In the course of this procedure of projection onto the $N$ bands,
all the relevant information, about the bands integrated away, was
encoded in the form of a gauge field, and appears in the EOM for the
wave packet as a Berry phase.

In order to illustrate this point,
let us investigate how those gauge fields
are expressed explicitly in terms of Bloch functions.
We will see in later sections that the concept of Bloch bands
is susceptible of perturbations varying incommensurately
to the lattice structure.
As a result Bloch electrons become subject to a (non-Abelian) 
gauge field in a $(2D+1)$-dimensional parameter space
$(\bar{\bf k},\bar{\bf x},t)$,
which we will call below the {\it reciprocal} space,
from the view point that it is a generalization
of the space spanned by $\bar{\bf k}$, the mean crystal
momentum of the wave packet.
The reciprocal vector potential takes the form of a $N\times N$ matrix,
whose elements are given by
\be
\big({\cal A}_q\big)_{mn}=
i\l\la u_{m} (\bar{\bf k},\bar{\bf x},t)\bigg|
{\rd u_n (\bar{\bf k},\bar{\bf x},t)\over\rd q}\r\ra
\label{vec}
\ee
where $q$ should be understood as a general coordinate
$q=\bar{k}_\mu, \bar{x}_\nu, t$ and $\mu,\nu=1,\cdots,D$.
$\bar{k},\bar{x}$ are center of mass coordinates defined in more 
precise terms, respectively, in Eqs. (\ref{kbar}) and (\ref{xbar}). 
$|u_n ({\bf k},\bar{\bf x},t)\ra=\exp(-i{\bf k}\cdot{\bf x})
|\phi_n ({\bf k},\bar{\bf x},t)\ra$ 
is the periodic part of a {\it local} 
\fn{The concept of {\it local} Bloch function will be
briefly introduced in Sec. 2.2 before being formulated in
more precise terms in Sec. 3.}
Bloch state; 
$\la{\bf x+a}|u_n({\bf k},\bar{\bf x})\ra=
\la{\bf x}|u_n({\bf k},\bar{\bf x})\ra$.
Inner products involving the periodic part
$u_n({\bf k},\bar{\bf x},t)\ra$,
mean an integration over the unit-cell,
with a normalization
$\la u_n({\bf k},\bar{\bf x},t)|u_n({\bf k},\bar{\bf x},t)\ra=1$.
In the abelian case $N=1$, this vector potential
is indeed related to the the reciprocal magnetic field
${\cal B}(\bar{\bf k})$ introduced in Eq. (\ref{recipro}) as
\[
{\cal B}(\bar{\bf k})=
{\rd\over \rd\bar{\bf k}}
\times {\cal A}_{\bar{\bf k}}.
\]
In the non abelian case, the gauge invariant
reciprocal field strength should be defined as
\be
{\cal F}_{q_1 q_2}=
\rd_{q_1}{\cal A}_{q_2}-\rd_{q_2}{\cal A}_{q_1}
+i\big[{\cal A}_{q_1},{\cal A}_{q_2}\big],
\label{field}
\ee
where $q_1,q_2=k_\mu, \bar{x}_\mu, t$.
Using a trivial relation
$\bla {\rd u_{m} \over\rd q}\big|u_n\bra
+\bla u_{m}\big|{\rd u_n \over\rd q}\bra=0$,
the last term of Eq. (\ref{field}) can be rewriten as
\[
i\sum_{l=1}^N\bigg(
\Bla {\rd u_{m} \over\rd q_2}\bigg|u_l\Bra
\Bla u_l\bigg|{\rd u_n\over\rd q_1}\Bra-
\Bla {\rd u_{m} \over\rd q_1}\bigg|u_l\Bra
\Bla u_l\bigg|{\rd u_n\over\rd q_2}\Bra\bigg),
\]
whereas,
\be
\Big(\rd_{q_1}{\cal A}_{q_2}-\rd_{q_2}{\cal A}_{q_1}\Big)_{mn}=
i\bigg(\l\la {\rd u_{m} \over\rd q_1}\bigg|{\rd u_n \over\rd q_2}\r\ra-
\l\la {\rd u_{m} \over\rd q_2}\bigg|{\rd u_n \over\rd q_1}\r\ra\bigg).
\label{berry}
\ee
Comparing those two equations, one can immediately see that
if $\sum_{l=1}^N \l.\big|u_l\r\ra\l\la u_l\big|\r.$ were 1, i.e.,
if $\{\l.\big|u_l\r\ra ; l=1,\cdots, N\}$ spanned a complete basis,
then ${\cal F}_{q_1 q_2}$ would vanish identically.
This indicates the fact that the nature of our gauge
field lies indeed in the projection of an available Hilbert space onto
the relevant $N$ bands. If $\l.\big|u_l\r\ra$ spanned a complete
basis, and no band were projected away, there would be
no information which should be encoded in the gauge fields.
Note also that Eq. (\ref{berry}) takes the familiar form of the
Berry curvature in the study of magnetic Bloch bands.
\cite{TKNN,CN}

\sb{Gauge field of two different origins}
\ind
The gauge field intorduced in Eq. (\ref{vec}) has two different
physical origins :
\ben
\item
Projection onto a subspace spanned by $N$ Bloch bands,
\item
Bloch basis moving in time.
\een
The first point has been already discussed in Sec. 2.1, whereas
the second point may need some explanation.
In the following sections, we will study the wave-packet dynamics
in the phase space in the presence of space and time dependent
external pertubation, which varies {\it incommensurately} to the
lattice structure.
In order to define a crystal momentum in such a situation,
we replace the spatial coordinate ${\bf x}$ in the perturbation
$\beta({\bf x},t)$, introduced as in Eq. (\ref{beta}), 
by the center-of-mass coordinate $\bar{\bf x}$ of a wave
packet under consideration.
This recovers the original lattice periodicity
of the Hamiltonian, leading us to the concept of 
{\it local} Hamiltonian (Eq. (\ref{Hloc})) and its {\it local}
Bloch eigenstates (Eq. (\ref{local})).
The above procedure is justified, whenever the external perturbation
varies sufficiently smoothly compared with the width of the
wave packet.
We then expand the wave packet in terms of the {\it local}
Bloch eigenstates, 
$|\phi_n({\bf k},\bar{\bf x}(t),t)\ra$,
which evolve as a function of time, both explicitly (through $t$) 
and implicitly (through $\bar{\bf x}(t)$). 
This is why our {\it local} Bloch function, or rather its periodic 
part, which has appeared in Eq. (\ref{vec}), depended not only
on ${\bf k}$ but also $\bar{\bf x}$ and $t$.
Because of the nature of our {\it local} Bloch basis,
such nontrivial gauge field structure as was introduced in 
the previous section emerges.
To be precise, we had better distinguish between
two different types of gauge field (strength) appearing in 
Eqs. (\ref{vec},\ref{field}):
(i) ${\cal F}_{\bar{k}_\mu \bar{k}_\nu}$,
(ii) ${\cal F}_{\bar{k}_\mu \bar{x}_\nu}$ and 
${\cal F}_{\bar{k}_\mu t}$.
Although the reciprocal field strength
introduced in Eq. (\ref{field}) has various components,
i.e., not only (a) ${\cal F}_{\bar{k}_\mu \bar{k}_\nu}$,
${\cal F}_{\bar{k}_\mu \bar{x}_\nu}$ and ${\cal F}_{\bar{k}_\mu t}$
((a)$=$(i)$+$(ii)),
but also 
(b) ${\cal F}_{\bar{x}_\mu \bar{x}_\nu}$ and
${\cal F}_{\bar{x}_\mu t}$,
the latter components (b) do not appear in our
EOM for the wave packet,
showing a clear contrast with Ref. \cite{SN}.
However, we will be working in the framework of a linear response 
theory, and within that framework our EOM turn out to be 
consistent, when $N=1$, with those of Ref. \cite{SN}.
This point will  be further clarified in Sec. 4.4 by performing
a simple power counting analysis.

We will see in detail in Sec. 4 that the two types of gauge field
\ben
\item
${\cal F}_{\bar{k}_\mu \bar{k}_\nu}$,
\item
${\cal F}_{\bar{k}_\mu \bar{x}_\nu}$ and
${\cal F}_{\bar{k}_\mu t}$.
\een
have actually slightly
different origins, as well as their different physical consequences
which we will discuss in Sec 5.
The former,
${\cal F}_{\bar{k}_\mu \bar{k}_\nu}$,
is indeed related to the projection of available Hilbert space
onto the relevant degenerate $N$ bands.
It yields a finite anomalous velocity, and plays a central role
in the understanding of AHE.
It appears in the presence of magnetic Bloch
bands and ferromagnetic backgrounds.
\cite{CN,OMN,RS1,JNM}.
On the other hand, the latter,
${\cal F}_{\bar{k}_\mu \bar{x}_\nu}$ and
${\cal F}_{\bar{k}_\mu t}$
appear only in the presence of the
{\it time-dependent Bloch basis}
mentioned above.

\sa{Statement of the problem}
\ind
Before discussing the EOM in the following section,
let us define and formulate our problem here as well as
listing all the assumptions we will make.
We stress here that all the approximation which we will make
are stated here, and that the derivation of the EOM in the
following section is indeed {\it exact}
under the assumptions made in this section.

Let us consider the motion of a wave packet of Bloch functions
under perturbations slowly varying in space and time.
This perturbation can be, e.g., external electro-magnetic field,
as was the case in the study of magnetic Bloch bands.
\cite{TKNN,CN}
The external perturbation $\beta ({\bf x},t)$, varing incommensurately
to the crystal structure, breaks the translational symmetry 
of the unperturbed Hamiltonian, 
\be
H_0({\bf p, x})={ {\bf p}^2\over 2m_e}+U({\bf x}),\ \ \
U({\bf x+a})=U({\bf x}).
\label{H0}
\ee
Eigenstates of the above Hamiltonian (\ref{H0}), 
i.e., Bloch bands (specified by band indices $n$) 
are characterized by crystal momenta ${\bf k}$,
\be
H_0\big|\phi_n^{(0)}({\bf k})\bra=\epsilon_n^{(0)}({\bf k})
\big|\phi_n^{(0)}({\bf k})\bra.
\label{en0}
\ee
Once the perturbation $\beta ({\bf x}, t)$ is switched on,
this crystal momentum ${\bf k}$ is no longer a good quantum number
of the system.
However, the typical wave length of the external perturbation
is longer by several order of magnitudes than the lattice constants,
in a physically relevant parameter regime of our interest.
In that case, intermediate length scales do exist,
to which our wave packet will belong,
in which the external perturbation $\beta ({\bf x}, t)$
can be regarded spatially constant at the zero-th order
of appoximation.
We are thus entitled to consider a wave packet,
well localized in this length scale of external perturbation,
which has also a peak sharp enough in the space of crystal momentum,
moving under pertubations slowly varying in space and time.

Let us now consider a wave packet, $\l.\big|\Psi(t)\r\ra$,
localized in the {\it phase space},
spanned by the real space
coordinate ${\bf x}$ and the crystal momentum ${\bf k}$,
in the vicinity of $(\bar{\bf k},\bar{\bf x})$.
For simplicity, and without losing generality,
we can assume that the wave packet has a symmetric and
smooth shape such that it has a well-distinguished peak
at $\big(\bar{\bf k}(t),\bar{\bf x}(t)\big)$ in the phase space,
where $\bar{\bf k}(t)$ and $\bar{\bf x}(t)$ should
coincide with the expectation value of ${\bf k}$ and ${\bf x}$
at a given time $t$.

Our present goal is to study, as accurately as possible, the quantum
mechanical motion of this wave packet, and derive the
effective equations of motion for $\bar{\bf x}$ and $\bar{\bf k}$.
As will soon become clearer, an interpretation in terms of
reciprocal gauge field (strength)
uncover the nature of various physical phenomena,
such as anomalous Hall effect (AHE),
\cite{CN,OMN,RS1,JNM}, spin Hall effect, \cite{MNZ}
and quantum charge/spin pumping. \cite{TN,RS2}

\sb{Assumption of slowly varying perturbation $\beta({\bf x}, t)$
- concept of the {\it local} Hamiltonian and its {\it local} Bloch bands}
\ind
We consider from now on a perturbation $\beta({\bf x}, t)$ introduced
in Eq. (\ref{beta}).
As far as the intermediate length scales discussed at the
beginning of this section are concerned,
$\beta({\bf x},t)$ can be regarded,
over the spread of our wave packet,
almost spatially constant.
We, therefore, choose, as the starting point of our analysis,
a Hamiltonian, dubbed in Ref. \cite{SN} as a {\it local} Hamiltonian,
in which ${\bf x}$-dependence of $\beta({\bf x},t)$ is replaced
by $\bar{\bf x}$, a constant at a given time :
\be
H_{\rm loc}=H({\bf p, x}; \beta(\bar{\bf x},t)).
\label{Hloc}
\ee
This $H_{\rm loc}$ has a very remarkable property ; 
at a given time $t$ it has the same translational symmetry 
as the non-perturbed Hamiltonian
$H_0=H({\bf p, x}; \beta=0)$, i.e., in other words,
$H_{\rm loc}$ can be diagonalized by a set of 
{\it local} Bloch eigenstates
$|\phi_n({\bf k},\bar{\bf x},t)\ra$ forming a
{\it local} band $\epsilon_n({\bf k},\bar{\bf x},t)$,
which now depends on $\bar{\bf x}(t)$ and $t$ :
\bea
H_{\rm loc}\l.|\phi_n({\bf k},\bar{\bf x},t)\r\ra=
\epsilon_n({\bf k},\bar{\bf x},t)
\l.|\phi_n({\bf k},\bar{\bf x},t)\r\ra.
\label{local}
\eea
We are actually considering a degenerate case where 
$\epsilon_n({\bf k},\bar{\bf x},t)\ (n=1,\cdots,N)$
takes the same value, which we define to be
$\epsilon_{\rm loc}({\bf k},\bar{\bf x},t)$, i.e.,
\be
\epsilon_{\rm loc}({\bf k},\bar{\bf x},t)\equiv
\epsilon_1({\bf k},\bar{\bf x},t)=\cdots=
\epsilon_N({\bf k},\bar{\bf x},t)
\label{eloc}
\ee
We will see below that
the concept of the {\it local} Hamiltonian and
its assoicated conduction bands plays a central role
in the derivation of EOM.

\sb{Construction of a wave packet}
\ind
Superposing {\it local} Bloch functions introduced above,
we now construct our wave packet.
In the spirit of Boltzmann's transport theory,
an exchange of energy between the electron and the environment
occurs only through scattering events.
In the following, we will investigate an adiabatic motion
of this wave packet. This picture should be valid over
the typical length scale of an {\it adiabatic} flight between two
scattering events, i.e., over the mean free path of an electron.
Let us now proceed step by step, making each logical step as
clear as possible.
\ben
\item
Let us first focus on the real space, in which
the electron wave packet is localized around $\bar{\bf x}$.
Then we can compose a wave packet
\fn{As has been discussed at the beginning of this section,
the expansion (\ref{Psi}) is justified, as far as the spread of
wave packet in real space is sufficiently small
compared with the typical length scale
over which the external perturbation can be regarded
almost constant.}
out of {\it local} Bloch functions associated with
the local Hamiltonian at $\bar{\bf x}$ :
\be
\l.|\Psi(t)\r\ra=\sum_{n=1}^N
\int d{\bf k} a_n({\bf k},t)\l.|\phi_n({\bf k},\bar{\bf x},t)\r\ra.
\label{Psi}
\ee
$a_n({\bf k},t)$ should be normalized properly.
The $\bar{\bf x}$-dependence of $\l.\big|\Psi(t)\r\ra$ is implicit
on the left hand side of Eq. (\ref{Psi}),
which is actually due to the time dependent Bloch basis,
$\l.|\phi_n({\bf k},\bar{\bf x},t)\r\ra$.
\item
In order for the self-consistency, we require that our wave
packet (\ref{Psi}) does give, the {\it correct} expectation value
of ${\bf x}$, i.e.,$\bar{\bf x}(t)=(\bar{x}_1(t),\cdots,\bar{x}_{D}(t))$:
\be
\bar{x}_\mu(t)=\l\la\Psi(t)\big| x_\mu \big|\Psi(t)\r\ra.
\label{xbar}
\ee
This guarantees that our wave packet yield, indeed, the
center-of-mass position {\it preassigned} in Eq. (\ref{Hloc}), 
and that our program makes a self-consistent closed loop.
\een
Our wave packet (\ref{Psi}) can be also regarded as a {\it functional} of
$a_n({\bf k},t)$, i.e., $|\Psi(t)\ra=|\Psi(\{a_{n}({\bf
k},t)\})\ra$,  in
which the coefficients $a_n({\bf k},t)$ are chosen
so that the self-consistency condition (\ref{xbar}) should be satisfied.
Eq. (\ref{xbar}) is, however, nothing but a weak constraint
compared with a huge number of degrees of freedom allowed for
$a_n({\bf k},t)$.
In order to specify with further precision the coefficients
$a_n({\bf k},t)$, we now turn our eyes to the ${\bf k}$-space.
As has been discussed at the beginning of this section,
we can consider, in the length scale of our interest, a wave packet
which is localized both in ${\bf x}$ and in ${\bf k}$.
We therefore require, in addition to Eq. (\ref{xbar}),
that our wave packet should {\it also} give the correct expectation
value of $k_\mu$ :
\be
\bar{k}_\mu(t)=
\l\la\Psi (t)\big| k_\mu \big|\Psi(t)\r\ra.
\label{kbar}
\ee
The $\bar{\bf k}$ dependence of $\l.\big|\Psi(t)\r\ra$
is thus encoded in $a_n({\bf k},t)$.
\fn{
So far, our treatment of ${\bf x}$ and ${\bf k}$ has not been
symmetric. This is entirely due to the fact that our perturbation
$\beta({\bf x},t)$ does not depend on ${\bf k}$.
We can consider, in principle and without much difficulty,
such a perturbation that depends on ${\bf k}$, and perform symmetric
treatment of ${\bf x}$ and ${\bf k}$. However, in this paper,
we restricted ourselves, for the clarity of the paper, to the former case.}

In the following section, we will derive the EOM for
$\bar{\bf x}(t)$ and $\bar{\bf k}(t)$,
following the quantum mechanical motion of the wave packet we have just
prepared.
In order to make the set of EOM self-contained, however, we need also
to take care of the motion of internal pseudospin degrees of freedom
spanned by $N$ bands.
For that purpose it will turn out to be convenient to separate
$a_{n}({\bf k},t)$ into its "phase" or pseudospin part,
${\bf z}^t({\bf k},t)=(z_1({\bf k},t),\cdots, z_N({\bf k},t))$,
and its "amplitude" part, $\rho(k,t)$, by introducing
\bea
a_n({\bf k},t)&=&\sqrt{\rho ({\bf k},t)}z_n({\bf k},t),
\nn \\
\rho({\bf k},t)&=&\sum_{n=1}^N \Big|a_n({\bf k},t)\Big|^2.
\nn
\eea
$|z_n|^{2}$ clearly represents the probability that
the electron wave packet sits on the $n$-th band among the
$N$-fold degenerated bands.
Thereby it corresponds to the internal
degrees of freedom associated with the wave packet,
such as spin and/or orbital, while $\rho({\bf k},t)$
is the momentum distribution function for the wave packet.
Since we assumed that the wave packet is well-localized not only
in its real space but also in its reciprocal space,
we can assume without any loss of generality
the following reduction formula,
\bea
\int d{\bf k} f({\bf k},t)\rho({\bf k},t)
= f(\bar{\bf k}(t),t)
\label{presc},
\eea
for any sufficiently smooth function $f({\bf k},t)$.
This prescription will be used frequently at the final stage of
the derivation of EOM.

\sb{First order perturbation theory with respect to 
$\beta({\bf x},t)$: a linear response theory}
\ind
The wave packet introduced above should obey
the Schr\"odinger equation
\be
i{\rd\over\rd t}\l.|\Psi(t)\r\ra=H
\l.|\Psi(t)\r\ra.
\label{Sch}
\ee
As we have briefly seen in the Introduction, our eventual
objective is to apply the EOM to the framework of the Boltzmannn
transport theory, using formula such as Eq. (\ref{J}),
in order to describe phenomena including the anomalous Hall
effect, spin Hall effect and quantum pumping, etc.
For that purpose it is enough to consider
a linear response of the system, keeping only the terms
up to first order of
$\partial \beta(\bar{\bf x},t)/\partial\bar{\bf x}$ and
$\partial \beta(\bar{\bf x},t)/\partial t$.

In the case of the electromagnetic fields,
the perturbation $\beta(\bar{\bf x},t)$ is embodied by
a vector potential ${\bf A}(\bar{\bf x},t)$, and 
a scalar potential $A_0(\bar{\bf x})$;
the full Hamiltonian reads
$H_0({\bf x},{\bf p}+e{\bf A}({\bf x},t))-eA_0({\bf x})$.
Thereby, a linear response to applied electro-maganetic fields,
${\bf E}=-\partial A_0/\partial {\bf x}-
\partial {\bf A}/\partial t$, 
${\bf B}=\rot {\bf A}$
corresponds to the first order perturbation theory w.r.t.
$\beta(\bar{\bf x},t)$.

We expand the Hamiltonian in powers of ${\bf x}-\bar{\bf x}$ as
\be
H=H_{\rm loc}+{1\over 2}
\sum_{\mu=1}^{D}\l\{
({x_{\mu}-\bar{x}_{\mu} }){\rd H_{\rm loc}\over\rd\bar{x}_{\mu}}+
{\rd H_{\rm loc}\over\rd\bar{x}_{\mu}}(x_{\mu}-\bar{x}_{\mu})
\r\},
\label{H1}
\ee
The first order term on the r.h.s.
is written in a symmetrical way in order to keep the Hamiltonian 
to be hermitian.
In the following, based on Eq. (\ref{H1}) 
we develop a systematic perturbation theory
w.r.t. $\beta(\bar{\bf x},t)$.
In this paper, we focus on the linear response of the system, 
keeping only the terms upto first order in the expansion.
Our treatment is, therefore, self-concistent in the framework of
linear response theory.

\sa{Equations of motion}
\ind
In this section, we sketch the derivation of EOM,
paying particular attention to,
how the two different types of reciprocal field strength,
introduced in Sec. 2, appear in the EOM.
Before going into the details of the derivation of EOM,
let us remind you that there are two possible sources of
Berry curvature in the reciprocal parameter space :
\ben
\item
projection of available Hilbert space onto the degenerated $N$
Bloch bands,
\item
{\it local} Bloch basis changing gradually in the course of
 time.
\een
The time dependence of the local Bloch basis stems, not only from the
explicit $t$ dependence of the {\it local} Hamiltonian,
$H_{\rm loc}$,
but also from our self-consistent treatment of the problem,
where the {\it local} Hamiltonian depends on the center-of-mass position
of the electron wave packet through the external perturbation,
$\beta(\bar{\bf x}(t),t)$.

In many respects, our point of view is reminiscent of
the standard Ehrenfest's theorem of quantum mechanics :
the expectation value of an operator, such as ${\bf x}$ or ${\bf p}$,
obeys a classical EOM.
We actually follow the same type of procedure as
the derivation of the Ehrenfest's theorem, and in this sense
our EOM can be regarded as a {\it generalized Ehrenfest's theorem}
for Bloch electrons under perturbations varying slowly in space
and time. We will come back to this point later.

\sb{Preliminaries}
\ind
We investigate, in this section, time evolution
of the wave packet constructed in Eq. (\ref{Psi}).
We are interested, not only in its motion in the phase space,
but also in the motion of its internal spin/orbital degrees of freedom.
In Sec. 5 we will develop further analyses,
from the viewpoint of transport phenomena,
on the dynamics
associated with such internal degrees of freedom.
Having in mind applications of our formalism to those fields,
we formulate our equations as generally as possible.
More concretely, we consider the time evolution of
an arbitary observable, ${\cal O}$, or rather of its
expectation value,
\[
\bar{\cal O}(t)=\bla\Psi (t)\big| {\cal O} \big|\Psi(t)\bra.
\label{Obar1}
\]
Since we have adopted, for the sake of simplicity, the Schr\"odinger
picture, as seen in Eq. (\ref{Sch}),
the wave functions evolve in time, while observables are
time-independent.
We develop later more detailed analyses on the EOM
focusing on the case where ${\cal O}=x_\mu$ or $k_\mu$,
but we consider a general observable ${\cal O}$
as far as possible in formulating our equations.
This will make it easier to apply our formalism to
further studies on the dynamics associated with
the internal degrees of freedom of Bloch electron.

Having those in mind, let us consider the expectation value
of an arbitrary observable ${\cal O}$,
\be
\bar{\cal O}(t)=
\sum_{m,n=1}^N\int d{\bf k} d{\bf k}^{\prime}
a_{m}^*({\bf k}^{\prime},t)
\bla{\cal O}\bra_{mn}({\bf k}^{\prime},{\bf k})
a_n({\bf k},t),
\label{Obar2}
\ee
where we have introduced an abbreviated notation for the matrix
elements of an operator ${\cal O}$ evaluated in the restricted
subspace spanned by $N$ Bloch bands, i.e.,
\be
\bla {\cal O} \bra_{mn}({\bf k}^{\prime},{\bf k})=
\l\la\phi_m({\bf k}^{\prime},\bar{\bf x},t)\big| {\cal O}
\big|\phi_n({\bf k},\bar{\bf x},t)\r\ra.
\label{Omn}
\ee
Note that in this restricted Hilbert space
not only $k_{\mu}$, $x_\mu$ or $H$ but also $\rdt$ are
considered to be an operator ${\cal O}$.
$\la {\cal O}\ra_{mn}(\bf{k',k})$ is generally
a $N\times N$ matrix for given $({\bf k}^{\prime},{\bf k})$,
whereas for a given $(m,n)$,
it has, in general, off-diagonal matrix elements, and
can be also regarded as a matrix in ${\bf k}$-space.
The presence of finite off-diagonal matrix elements
of  $\la {\cal O}\ra_{mn}(\bf{k',k})$,
either in the $k$ space or in the pseudospin space
prevents some observables from commuting each other,
thereby makes them non-abelian.

Let us first consider two concrete examples :
\ben
\item
Case of ${\cal O}=x_\mu$:
The matrix elements of an observable $x_\mu$ are
\be
\bla x_\mu\bra_{mn}({\bf k',k})=
i\delta({\bf k}^{\prime}-{\bf k})\delta_{mn}{\rd\over\rd k_\mu}+
\delta({\bf k}^{\prime}-{\bf k})\Big({\cal A}_{k_\mu}\Big)_{mn}
\label{xmn1}
\ee
The first term is off-diagonal in ${\bf k}$-space, when ${\bf k}$ is
discrete,
due to the ${\bf k}$-derivative, but is diagonal w.r.t. the band index.
In Eq. (\ref{xmn1}), we kept both ${\bf k'}$ and ${\bf k}$ indices
in order to emphasize the fact that this first term is off-diagonal.
In the following, we will omit quite frequently the 
${\bf k'}$-index, 
pretending that $\bla x_\mu\bra_{mn}({\bf k',k})$
is diagonal in ${\bf k}$-space
after $\delta ({\bf k'}-{\bf k})$ is intergrated away.
On the contrary, the second term is diagonal in ${\bf k}$-space,
but the reciprocal vector potential ${\cal A}_{\bf k}$
defined similarly to Eq. (\ref{vec}), as,
\[
\Big({\cal A}_{k_\mu}\Big)_{mn}=
i\l\la u_m ({\bf k},\bar{\bf x},t)\Big|
{\rd u_n ({\bf k},\bar{\bf x},t)\over\rd k_\mu}\r\ra,
\]
has off-diagonal matrix element between different bands.
In the above equations we did not write down the explicit $t$-dependence
of $\bar{\bf x}(t)$ in the brackets.
\item
Case of ${\cal O}=k_\mu$:
This case is even simpler.
The crystal momentum ${\bf k}$ is diagonal both in ${\bf k}$
and in pseudospin indices,
\be
\bla k_\mu\bra_{mn}({\bf k',k})=
\delta({\bf k}^{\prime}-{\bf k})\delta_{mn}k_{\mu}.
\label{kmn}
\ee
\een
Let us further investigate the {\it off-diagonal} components
of $\bla x_\mu\bra_{mn}({\bf k',k})$.
We focus here on its commutation relation in ${\bf k}$-space.
Since the first term on the r.h.s. of Eq. (\ref{xmn1}) is off-diagonal
and the second term is not proproportional to an identity matrix,
these two terms do not commute each other.
One can indeed verify
\be
\Big[\bla x_\mu\bra, \bla x_\nu\bra\Big]_{mn}({\bf k})=
i{\cal F}_{k_\mu k_\nu},
\label{xx}
\ee
where $[A,B]=AB-BA$ is a standard commutator of 
two $N\times N$ matrices $A,B$.
Thus the noncommutativity of
$\bla x_\mu\bra_{mn}({\bf k})$
turns out to be the origin of the emergence of
${\cal F}_{k_\mu k_\nu}$.
Another important remark on the Eq. (\ref{xmn1}) is that
it leads us to introduce naturally the concept of
{\it covariant} derivative in momentum space
\cite{MNZ}, defined as
\be
\Big(\nabla_{k_\mu}\Big)_{mn}=
\delta_{mn}{\rd\over\rd k_\mu}-
i\Big({\cal A}_{k_\mu}\Big)_{mn}.
\label{cov}
\ee
In terms of the covariant derivative $\big( \nabla_{k_\mu}\big)_{mn}$
thus introduced, the matrix elements $\bla x_\mu\bra_{mn}({\bf k})$ 
can be rewritten simply as
\be
\bla x_\mu\bra_{mn}({\bf k})=i\Big(\nabla_{k_\mu}\Big)_{mn}.
\label{xmn2}
\ee
The commutator between two covariant derivatives along 
different axes is directly related a non-abelian Berry 
curvature ;
\be
\Big[ \nabla_{k_\mu}, \nabla_{k_\nu} \Big]_{mn}=
-i{\cal F}_{k_\mu k_\nu}.
\label{dd}
\ee
In geometric terms, Eq. (\ref{dd}) can be interpreted
in such a way that two parallel transports along different axes
on a curved surface generally do not commute each other.

\sb{To derive the EOM}
Let us now consider the time derivative of the
expectation value, $\bar{\cal O}(t)$.
Expanding $\l|\Psi(t)\r\ra$ in terms of the local Bloch functions
as Eq. (\ref{Psi}), one can classify the time derivative of
$\bar{\cal O}(t)$ into three parts :
\bea
{d\bar{\cal O}(t)\over dt}&=&\sum_{\sigma,\sigma^{\prime\prime}} 
\int d{\bf k}d{\bf k}^{\prime\prime}
{\rd a_{\sigma}^{\ast}({\bf k},t)\over\rd t}
\bla {\cal O} \bra_{\sigma\sigma^{\prime\prime}}
({\bf k},{\bf k}^{\prime\prime})
a_{\sigma^{\prime\prime}}({\bf k}^{\prime\prime},t)
\nn \\
&+&
\sum_{\sigma^{\prime}\sigma^{\prime\prime}}
\int d{\bf k}^{\prime}d{\bf k}^{\prime\prime}
a_{\sigma^{\prime}}^{\ast}({\bf k}^{\prime},t)
\Big(\rdt
\bla {\cal O} \bra_{\sigma^{\prime}\sigma^{\prime\prime}}
({\bf k}^{\prime},{\bf k}^{\prime\prime})
\Big)a_{\sigma^{\prime\prime}}({\bf k}^{\prime\prime},t)
\nn \\
&+&
\sum_{\sigma^{\prime}\sigma}
\int d{\bf k}^{\prime}d{\bf k}
a_{\sigma^{\prime}}^{\ast}
({\bf k}^{\prime},t)
\bla {\cal O} \bra_{\sigma^{\prime}\sigma}
({\bf k}^{\prime},{\bf k})
{\rd a_{\sigma}({\bf k},t)\over\rd t}.
\label{dodt}
\eea
We have in mind that the operator
${\cal O}$ is either $x_{\mu}$, $k_{\mu}$ or some other
observables.
In the case of standard Ehrenfest's theorem,
\[
\ddt\l\la\psi\big|{\cal O}|\psi\r\ra=
i\l\la\psi\big|[H,{\cal O}]\big|\psi\r\ra,
\]
the second term of Eq. (\ref{dodt}) does not exist,
since the matrix element,
$\bla {\cal O} \bra_{mn}({\bf k}^{\prime},{\bf k})$
is time-dependent only when the local Bloch basis evolves in time.
The first and the third terms, i.e., the change of
expansion coefficients $a_{\sigma}({\bf k},t)$
yields a commutator, $[H,{\cal O}]$.
They contain, however, also a Berry connection contribution,
which, together with the second term, produce a new type
of contribution, which we will call $\Omega_{\cal O}^{(2)}$ in Eq.
(\ref{dec}).
The first term of Eq. (\ref{dec}), $\Omega_{\cal O}^{(1)}$, is a
generalizaton
of the standard Ehrenfest's commutator $[H,{\cal O}]$, 
which induces, when ${\cal O}=x_\mu$,
${\cal F}_{\bar{k}_\mu\bar{k}_\nu}$in the EOM.
On the other hand, the second term, $\Omega_{x_\mu}^{(2)}$, can be
rewritten in terms of ${\cal F}_{\bar{k}_\mu\bar{x}_\nu}$
and ${\cal F}_{\bar{k}_\mu t}$.

In order to rewrite the Eq. (\ref{dodt}) in terms of
$\Omega_{\cal O}^{(1)}$ and $\Omega_{\cal O}^{(2)}$,
let us first look into the following relation,
\be
{\rd a^{\ast}_n ({\bf k},t)\over \rd t}=
\sum_{m=1}^N
a^{\ast}_m({\bf k},t) \Bigg[
\Bla\rdt\Bra_{mn}+
i\bla H\bra_{mn}({\bf k})
\Bigg] .
\label{dadt}
\ee
The first term is a Berry connection contribution.
As is clear when it is written more precisely as
\be
\Bla\rdt\Bra_{mn}=
\Bla u_m \Big| {\rd u_n\over\rd t}\Bra=
\Bla u_m({\bf k},\bar{\bf x}(t),t)\Big|\rdt\Big|
u_n({\bf k},\bar{\bf x}(t),t)\Bra,
\ee
it emerged as a result of
the time evolution of the {\it local} Bloch basis.
On the other hand,
the second term of Eq. (\ref{dadt})
yields the commutation $[H, {\cal O}]_{mn}$
in Eq. (\ref{dec}).
Note also that the derivative $\rd/\rd t$ in Eq. (\ref{dadt})
picks up both the {\it explicit} and
{\it implicit} $t$-dependence.
Correspondingly, one can also rewrite the first term of
Eq. (\ref{dadt}) using two types of the gauge field introduced
in Sec. 2, i.e.,
\be
\Bla\rdt\Bra_{mn}=
\Big({\cal A}_{\bar{x}_\nu}\Big)_{mn}{d\bar{x}_\nu\over dt}+
\Big({\cal A}_{t}\Big)_{mn}.
\ee

Our next objective is to calculate the time derivative of
an operator such as $\bar{\bf k}, \bar{\bf x}$
and express them in such a way that their interpretation
in terms of the reciprocal field strength will become
as easy as possible.
For that purpose, we rearrange the terms in Eq. (\ref{dodt})
into two parts, $\Omega_{\cal O}^{(1)}$ and $\Omega_{\cal O}^{(2)}$
as,
\fn{In order to obtain Eq. (\ref{dec}),
one has only to substitute
{\it literally} Eqs. (\ref{dadt}) into the expression for
$d\bar{\cal O}/dt$ in Eqs. (\ref{dodt}), and rename the dummy
variables in the following way :
\bea
k^{\prime}&\gt&k_1,
k^{\prime\prime}\gt k_2,
k\gt k_3,
\nn \\
\sigma^{\prime}&\gt&m,
\sigma^{\prime\prime}\gt n,
\sigma\gt l.\nn
\eea}
\bea
\ddt{\bar{\cal O}}(t)&=&
\Omega_{\cal O}^{(1)}+\Omega_{\cal O}^{(2)},
\nn \\
\Omega_{\cal O}^{(1)}&=&
i\sum_{m,n=1}^N
\int d{\bf k_1} d{\bf k_2} a_{m}^{\ast}({\bf k_1})
\Big[\bla H\bra, \bla{\cal O}\bra\Big]_{mn}({\bf k_1},{\bf k_2})
a_n({\bf k_2}),
\nn \\
\Omega_{\cal O}^{(2)}&=&
\sum_{m,n=1}^N
\int d{\bf k_1} d{\bf k_2} a_{m}^{\ast}({\bf k_1})
\Bigg\{\Big[\Bla\rdt\Bra, \bla{\cal O}\bra \Big]_{mn}({\bf k_1},{\bf k_2})
\nn \\
&+&\Big(\rdt\bla {\cal O}\ra_{mn}({\bf k_1,k_2})\Big)
\Bigg\}
a_n({\bf k_2})
\label{dec}
\eea
In Eq. ({\ref{dec}) we did not write down explicitly, for the sake of
simplicity, the dependence on $\bar{\bf x}$  and the $t$ in
$\l.\big|\phi_n({\bf k})\r\ra=\l.\big|\phi_n({\bf k},\bar{\bf x}(t),t)\r\ra$.

As has been announced in advance,
$\Omega_{\cal O}^{(1)}$ is a generalization (or, rather a restricted
version)
of the standard Ehrenfest's commutator, coming exclusively from the
first and third terms of Eq. (\ref{dodt}), whereas
$\Omega_{\cal O}^{(2)}$ is a new type of contribution, which is
a collection of Berry curvature terms from all the three parts of
Eq. (\ref{dodt}).
Not only have they different origins, but also are they
susceptible of different physical interpretations in terms of the
reciprocal field strength.
We will see in Sec. 4.2 that in the particular case of ${\cal O}=x_\mu$,
the two contributions, $\Omega_{x_\mu}^{(1)}$ and $\Omega_{x_\mu}^{(2)}$
are related actually to the two different parts of the gauge field
introduced in Sec. 2, i.e.,
(i) ${\cal F}_{k_{\mu}k_{\nu}}$, and (ii) ${\cal F}_{k_\mu x_\mu}$,
${\cal F}_{k_{\mu}t}$.

Having in mind what has been stated above, we can now
derive the EOM for
$\bar{x}_{\mu}(t)$ and $\bar{k}_{\mu}(t)$.
Let us first consider the case of ${\cal O}=k_{\mu}$.
As seen in Eq. (\ref{kmn}), the momentum operator $k_{\mu}$ is not
only diagonal in ${\bf k}$ coordinates and band indices, but
also its matrix element is time-independent. Thus
only the first term $\Omega_{k_\mu}^{(1)}$ contributes to its EOM .
Futhermore, among various matrix elements of the Hamiltonian
given in (\ref{Hmn1},\ref{Hmn2}), only those terms which contain off-diagonal
matrix elements w.r.t. ${\bf k}$ indices contribute to its
commutator with $k_{\mu}$. As a result, its EOM turns out to be simplified
as,
\be
\frac{d\bar{k}_\mu(t)}{dt}=- \int d{\bf k}\rho({\bf k},t)
{\rd \epsilon_{\rm loc}({\bf k},\bar{\bf x}, t)\over\rd\bar{x}_\mu}=
-{\rd \epsilon_{\rm loc}(\bar{\bf k},\bar{\bf x}, t)\over\rd\bar{x}_\mu}.
\label{dkdt}
\ee
In the second equality, we replaced ${\bf k}$ in the integrant by
its mean value, following the prescription given in Eq. (\ref{presc}).
This is nothing but the standard EOM for the momentum of the electron
wave packet shown in Eq.(\ref{C1}).

As for the position operator $x_{\mu}$, Eq. (\ref{xmn1})
contains both the ${\bf k}$-derivative and
time-dependent matrix elements between different band indices.
As a result, the EOM for the real space coordinate
is subject to a drastic change in comparison
with Eq. (\ref{C2}).
In Sec. 2, we classified the reciprocal fields into two categories,
i.e., (i) ${\cal F}_{k_{\mu}k_{\nu}}$, and
(ii) ${\cal F}_{k_{\mu}t}$ and
${\cal F}_{k_{\mu}x_{\nu}}{d\bar{x}_{\nu}\over dt}$.
We will see in the next section that the decomposition (\ref{dec})
clearly demonstrates why we classified them in that way.
We have studied on a very general basis in this section that
the two components,
$\Omega_{\cal O}^{(1)}$ and $\Omega_{\cal O}^{(2)}$,
are structurally well distinguishable, and have completely
different nature.
We will see more specifically in the next section that
$\Omega_{x_\mu}^{(1)}$ and $\Omega_{x_\mu}^{(2)}$
are realted respectively to the reciprocal fields (i) and (ii).
Thus different origins of two types of reciprocal fields
will be uncovered.
The fact that the classification of reciprocal fields
discussed in Sec. 2 can be done explicitly and unambiguously
as the decomposition (\ref{dec}),
is actually one of the main advantages of our approach.
Let us now turn to a close inspection of the nature of
$\Omega_{x_\mu}^{(1)}$ and $\Omega_{x_\mu}^{(2)}$.

\sb{Nature of $\Omega_{x_\mu}^{(1)}$ and $\Omega_{x_\mu}^{(2)}$}

\ind
In this section let us further analyze the nature of decopmposition
(\ref{dec}), focusing on the case of ${\cal O}=x_{\mu}$.
$\bar{\bf x}(t)$ is given by Eq. (\ref{xbar}) together with
Eq. (\ref{xmn1}).
The first term of Eq. (\ref{dec}),
$\Omega^{(1)}_{x_\mu}$ in the present case, has particularly a familiar
form,
which often appears in the context of the {\it Ehrenfest's theorem}
\be
\ddt\l\la\psi\big|x_\mu|\psi\r\ra=
i\l\la\psi\big|[H,x_\mu]\big|\psi\r\ra=
\frac{\l\la\psi\big|p_{\mu}\big|\psi\r\ra}{m}.
\label{QM}
\ee
The similarity between $\Omega_{x_{\mu}}^{(1)}$
and Eq. (\ref{QM}) becomes clearer, when one expands
the wave packet $\l.|\psi\r\ra$ in terms of a complete set of bases
$\l.|\phi_\alpha\r\ra$ as
$\l.|\psi\r\ra=\sum_\alpha a_\alpha\l.|\phi_\alpha\r\ra$.
The difference is that the set of bases used in the expansion
was {\it complete} in Eq. (\ref{QM}), whereas it was
{\it restricted to $N$ Bloch bands} in $\Omega_{x_{\mu}}^{(1)}$.
This constraint is the origin of non-vanishing field strengths.

Let us now proceed to rewrite $\Omega_{x_{\mu}}^{(1)}$
in terms of the reciprocal field strength,
${\cal F}_{k_\mu k_\nu}$, defined in (\ref{field}).
The matrix elements of the Hamiltonian, i.e., Eq. (\ref{H1})
in the restricted subspace, spanned by $N$ Bloch bands
are calculated to be
\bea
\bla H\bra_{mn}({\bf k}^{\prime},{\bf k})&=&
\bla\phi_{m}({\bf k}^{\prime},\bar{\bf x},t)\big|
H\big|\phi_n({\bf k},\bar{\bf x},t)\bra=
\delta({\bf k}^{\prime}-{\bf k})\bla H\bra_{mn}({\bf k})
\label{Hmn1}
\\
\bla H\bra_{mn}({\bf k})&=&\epsilon_{\rm eff}-
{\rd \epsilon_{\rm loc}\over\rd\bar{x}_\nu}\bar{x}_{\nu}
+\frac{i}{2}\bigg\{
{\rd \epsilon_{\rm loc}\over\rd\bar{x}_\nu}
\big( \nabla_{k_{\nu}} \big)_{mn} +
\big( \nabla_{k_{\nu}} \big)_{mn}
{\rd \epsilon_{\rm loc}\over\rd\bar{x}_\nu}
\bigg\},
\label{Hmn2}
\eea
where 
$\epsilon_{\rm loc}=\epsilon_{\rm loc}({\bf k},\bar{\bf x},t)$ is a degenerate
eigenvalue of the local Hamiltonian
\fn{See Eqs. (\ref{Hloc},\ref{eloc}). Recall also that
\[
\bla H_{\rm loc}\bra_{mn}({\bf k}^{\prime},{\bf k})=
\bla\phi_{m}({\bf k}^{\prime},\bar{\bf x},t)\big|
H_{\rm loc}\big|\phi_n({\bf k},\bar{\bf x},t)\bra=
\delta({\bf k}^{\prime}-{\bf k})\delta_{mn}\epsilon_{\rm loc}({\bf k},\bar{\bf x},t)
\]
is proportional to an identity in the pseudospin space.}.
We also introduced a renormalized energy,
$\epsilon_{\rm eff}({\bf k},\bar{\bf x},t)$,
which takes the form of a $N$ by $N$ matrix whose
$(m,n)$-components are given by
\be
\epsilon^{\rm eff}_{mn}({\bf k},\bar{\bf x},t)=
\epsilon_{\rm loc}({\bf k},\bar{\bf x},t)\delta_{mn}+
\Delta\epsilon_{mn}({\bf k},\bar{\bf x},t).
\label{eff}
\ee
Its off-diagonal matrix elements are due to correction terms,
\bea
\Delta\epsilon_{mn}({\bf k},\bar{\bf x},t)&=&
\frac{i}{2}
\l\la {\rd u_{m}({\bf k},\bar{\bf x},t)\over\rd k_\mu}
\Big|
\Big(H_{\rm loc}-\epsilon_{\rm loc}({\bf k},\bar{\bf x},t)\Big)
\Big|
{\rd u_n({\bf k},\bar{\bf x},t) \over\rd \bar{x}_\mu}\r\ra
\nn \\
&-&\frac{i}{2}
\l\la {\rd u_{m}({\bf k},\bar{\bf x},t)\over
\rd \bar{x}_\mu}
\Big|
\Big(H_{\rm loc}-\epsilon_{\rm loc}({\bf k},\bar{\bf x},t)\Big)
\Big|
{\rd u_n({\bf k},\bar{\bf x},t) \over\rd k_\mu}\r\ra,
\nn
\eea
where the summation over $\mu=1,\cdots,D$ was assumed implicitly. 
Using the matrix elements given in
Eqs. (\ref{Hmn1},\ref{Hmn2}) one can rewrite $\Omega_{x_{\mu}}^{(1)}$
in the following way,
\fn{Details are given in Appendix A.} 
\bea
\Omega_{x_{\mu}}^{(1)}&=&\sum_{m,n=1}^N\int d{\bf k} \rho({\bf k},t) 
z_{m}^{\ast}({\bf k},t)\bigg\{\Big[\nabla_{k_{\mu}}, 
\epsilon_{\rm eff}({\bf k},\bar{\bf x},t)\Big]_{mn}
\nn \\
&+&\sum_{\nu=1}^{D}\Big({\cal F}_{k_\mu k_\nu}\Big)_{mn}
{\rd \epsilon_{\rm loc}({\bf k},\bar{\bf x},t)\over \rd \bar{x}_\nu}
\bigg\}z_n({\bf k},t)+
\Delta \Omega^{(1)}_{x_{\mu}},
\label{omega1a}
\eea
Apart from the energy correction $\Delta \epsilon$,
the first term of $\Omega_{x_{\mu}}^{(1)}$
is nothing but a standard velocity term, i.e., the first term in
the r.h.s. of Eq. (\ref{C2}).
A remark worth mentioning here is that
the covariant derivative in the commutator plays a central
role in ensuring the $SU(N)$ gauge invariance of final results,
which we will see later.
$\Delta \Omega^{(1)}_{x_{\mu}}$ is a irrelevant term
\fn{Its explicit form is given in Eq. (\ref{irr}) in 
Appendix A.}
which vanishes with the help of prescription introduced in
Eq. (\ref{presc}).
For a later convenience, let us introduce
the following abbreviated vector notation for
$z_n\big(\bar{\bf k}(t), t\big)$ :
\be
\bar{\bf z}(t)^\dagger\equiv
\Big(
z_1^* \big(\bar{\bf k}(t), t\big),\cdots,
z_N^* \big(\bar{\bf k}(t), t\big)
\Big),
\label{z}
\ee
where we always have in mind the prescription (\ref{presc}).
Using this notation,
one can further rewrite Eq. (\ref{omega1a}) as
\be
\Omega^{(1)}_{x_{\mu}}= \sum_{m,n=1}^{N} 
\bar{z}_{m}(t)^{\ast}\ \bigg\{
\Big[\nabla_{\bar{k}_{\mu}},\epsilon_{\rm eff}(\bar{\bf k})\Big]_{mn}+
\sum_{\nu=1}^{D}
\Big({\cal F}_{\bar{k}_{\mu}\bar{k}_{\nu}}\Big)_{mn}
\frac{\rd \epsilon_{\rm loc}(\bar{\bf k})}{\rd \bar{x}_{\nu}}\bigg\}
\bar{z}_{n}(t),
\label{omega1b}
\ee
where $\bar{\bf x}$-dependence of $\epsilon_{\rm loc}(\bar{\bf k})$
is not written explicitly.

In contrast to $\Omega_{x_\mu}^{(1)}$,
the second term of Eq. (\ref{dec}),
$\Omega_{x_\mu}^{(2)}$ in the present case,
would not have existed,
unless the local Bloch basis had evolved in time.
However, in a general situation described by a time-dependent Bloch basis,
there is no reason to believe that $\Omega_{x_\mu}^{(2)}$ should vanish.
Indeed, we will give you in Sec. 5 some concrete examples
where a finite contribution from $\Omega_{x_\mu}^{(2)}$ plays a crucial
role in determining the physical properties of the system.
Using the Eqs. (\ref{dadt}), one can easily verify that
$\Omega_{x_\mu}^{(2)}$ are related to the second category of reciprocal
fields, i.e.,
${\cal F}_{k_{\mu}t}$ and
${\cal F}_{k_{\mu}x_{\nu}}{d\bar{x}_{\nu}\over dt}$.
$\Omega_{x_\mu}^{(2)}$ can be rewritten as
\bea
\Omega_{x_\mu}^{(2)}&=&-
\sum_{m,n=1}^N \int d{\bf k} 
\rho({\bf k},t)z_{m}^{\ast}({\bf k},t) 
\bigg\{\Big({\cal F}_{k_\mu \bar{x}_\nu}\Big)_{mn}
{d\bar{x}_\nu\over dt}+\Big({\cal F}_{k_\mu t}\Big)_{mn}
\bigg\}z_n({\bf k},t)
\nn \\
&=&-\bar{\bf z}(t)^{\dagger}  
\bigg\{\sum_{\nu=1}^{D}{\cal F}_{\bar{k}_\mu \bar{x}_\nu}
{d\bar{x}_\nu\over dt}+{\cal F}_{\bar{k}_\mu t}
\bigg\} \bar{\bf z}(t),
\label{omega2}
\eea
where the summation over $\nu=1,\cdots,D$ was omitted in the first line.
The decomposition (\ref{dec}) together with
Eqs. (\ref{omega1b}, \ref{omega2}) gives a complete physical
justification of the classification of ${\cal F}_{q_1 q_2}$
done in Sec. 2.
In other approaches \cite{SN,KT}
the two types of reciprocal fields appear 
in an indistinguishable manner,
and two different origins of reciprocal gauge field studied
in this paper remain to be hidden.

\sb{$SU(N)$ gauge invariance}
\ind
We have successfully related the two contributions
to $\ddt \bar{\bf x}(t)$ in the decomposition (\ref{dec}),
i.e., $\Omega_{x_\mu}^{(1)}$ and $\Omega_{x_\mu}^{(2)}$,
respectively, to two types of gauge invariant reciprocal fields,
(i) ${\cal F}_{k_{\mu}k_{\nu}}$, and
(ii) ${\cal F}_{k_{\mu}t}$ and
${\cal F}_{k_{\mu}x_{\nu}}{d\bar{x}_{\nu}\over dt}$.
Together with Eq. (\ref{dkdt}), this allows us to rewrite our EOM
for $\bar{\bf x}(t)$ and $\bar{\bf k}(t)$ as 
\bea
{d\bar{x}_\mu\over dt}&=&\bar{\bf z}^\dagger
\bigg\{\Big[\nabla_{\bar{k}_{\mu}},\epsilon_{\rm eff}\Big]-
{\cal F}_{\bar{k}_\mu \bar{k}_\nu}{d\bar{k}_\nu\over dt}-
{\cal F}_{\bar{k}_\mu \bar{x}_\nu}{d\bar{x}_\nu\over dt}-
{\cal F}_{\bar{k}_\mu t}
\bigg\} \bar{\bf z}, 
\label{EOM1} \\
{d\bar{k}_\mu\over dt}&=&-
{\rd \epsilon_{\rm loc}(\bar{\bf k},\bar{\bf x},t)\over\rd\bar{x}_\mu}.
\label{EOM2}
\eea
Repeated indices $\nu$ should be summed over $\nu=1,\cdots,D$.
The {\it effective} energy $\epsilon_{\rm eff}$ is related 
to the local $\epsilon_{\rm loc}$ as Eq. (\ref{eff}).
In Eqs. (\ref{EOM1},\ref{EOM2}),
$\epsilon_{\rm eff}$ and $\epsilon_{\rm loc}$ are
functions of $\bar{\bf k},\bar{\bf x}, t$, i.e.,
$\epsilon_{\rm eff}=
\epsilon_{\rm eff}(\bar{\bf k},\bar{\bf x},t)$,
$\epsilon_{\rm loc}=
\epsilon_{\rm loc}(\bar{\bf k},\bar{\bf x},t)$.
Below let us verify explicitly that Eqs. (\ref{EOM1})
are indeed $SU(N)$ gauge invariant.
In order to obtain a complete set of EOM, however,
we still need to know an EOM for $\bar{\bf z}(t)$ defined
in (\ref{z}).
The details of its derivation is given in Appendix B, and the
result is,
\be
i\frac{d \bar{\bf z}}{dt}= \Bigg(\epsilon_{\rm eff}-
\bar{x}_{\mu}\frac{\rd \epsilon_{\rm loc}}
{\rd \bar{x}_{\mu}}{\bf 1}-
{d\bar{k}_\mu\over dt}{\cal A}_{\bar{k}_\mu}-
{d\bar{x}_\mu\over dt}{\cal A}_{\bar{x}_\mu}-
{\cal A}_{t}
\Bigg)\bar{\bf z},
\label{EOM3}
\ee
where repeated indices $\mu$ should be summed over 
$\mu=1,\cdots,N$, and again,
$\epsilon_{\rm eff}=
\epsilon_{\rm eff}(\bar{\bf k},\bar{\bf x},t)$,
$\epsilon_{\rm loc}=
\epsilon_{\rm loc}(\bar{\bf k},\bar{\bf x},t)$.
Eqs. (\ref{EOM1},\ref{EOM2},\ref{EOM3}) consitute
the central result of this paper together with their
interpretation in terms of the field strength ${\cal F}_{q_1 q_2}$,
which will be further discussed in Sec. 5.

Before turning to the discussion on gauge invariance,
let us make a few comments on Eq. (\ref{EOM3}).
On the r.h.s. the diagonal part of $\epsilon_{\rm eff}$, i.e.,
$\epsilon_{\rm loc}(\bar{\bf k}){\bf 1}$, tpgether with the second 
term simply give rise to a usual $U(1)$ phase factor associated
with an effective energy,
$\epsilon_{\rm loc}(\bar{\bf k},\bar{\bf x},t)-
\sum_{\mu=1}^{N}\bar{x}_{\mu}
\frac{\rd \epsilon_{\rm loc}(\bar{\bf k},\bar{\bf x},t)}
{\rd \bar{x}_{\mu}}$.
On the other hand,
$\Delta\epsilon({\bf k})$ generally has
off-diagonal matrix elements between different bands
and thereby yields a nontrivial $SU(N)$ phase factor,
which corresponds to the precession of the spin and/or orbital
associated with the wave packet.
The remaining terms of Eq. (\ref{EOM3}) represents 
a Berry-Wilczek-Zee phase \cite{WZ} 
originating from the adibatic motion of the wave packet.
The first two terms are due to its motion in
$(\bar{\bf k}$,$\bar{\bf x})$-space.
In the Abelian case ($N=1$), the EOM for $\bar{\bf x}(t)$
and $\bar{\bf k}(t)$, i.e., Eqs. (\ref{EOM1},\ref{EOM2}),  
are independent of the motion of phase degree of freedom, 
$\bar{\bf z}(t)$, whereas
$\bar{\bf z}(t)$ acquires a quantal phase due to the
evolution of $\bar{\bf x}$ and $\bar{\bf k}$ :
$\exp\big[i\int dt \big({d\bar{k}_\mu\over dt}{\cal A}_{\bar{k}_\mu}+
{d\bar{x}_\mu\over dt}{\cal A}_{\bar{x}_{\mu}}+{\cal A}_{t}
\big)\big]$ 
where the summation over $\mu$ was omitted.
This is analogous to the Berry quantal phase \cite{berry}.

Since the $N$-fold Bloch states are energetically
degenerate over the whole Brillouin zone, these EOM should
be independent of the choice of $N$ Bloch bases and be
invariant under the following gauge transformation :
\bea
\big|\tilde{u}_{n}\big(\bar{\bf k},\bar{\bf x},t\big)\bra
&=&\sum_{m=1}^{N}\big|u_{m}\big(\bar{\bf k},\bar{\bf x},t\big)\bra
g_{mn}\big(\bar{\bf k},\bar{\bf x},t\big)
\nn \\
\tilde{\bf z}(\bar{\bf k},t) &=&
g^{-1}\big(\bar{\bf k},\bar{\bf x},t\big)
{\bf z}(\bar{\bf k},t).
\label{GT}
\eea
Here $\big|u_n\big(\bar{\bf k},\bar{\bf x},t\big)\bra$ and
${\bf z}(\bar{\bf k},t)$ are transformed inversely to each
other, making the l.h.s. of Eq. (\ref{Psi}) invariant.
The gauge field and the field strength associated with it
are transformed in the following way,
\bea
\tilde{\cal A}_{q_{\mu}}&=& g^{-1}
{\cal A}_{q_\mu}g-
ig^{-1}\frac{\rd g}{\rd q_{\mu}}, 
\nn \\
\tilde{\cal F}_{q_{\mu}q_{\nu}}&=&g^{-1}
{\cal F}_{q_\mu q_\nu}g,
\nn 
\eea
where $q_\mu=\bar{\bf k},\bar{\bf x},t$.
Concomitantly, the covariant derivative, defined generally
for this $q_\mu$ as
\be
\nabla_{q_{\mu}}=\frac{\rd}{\rd q_{\mu}}- 
i{\cal A}_{q_{\mu}},
\ee
obeys the following transformation rule:
\be
\tilde{\nabla}_{q_{\mu}}=g^{-1}\nabla_{q_{\mu}}g.
\ee
Futhermore, one can also check that the $N\times N$ matrix
$\epsilon_{\rm eff} (\bar{\bf k})$ is transformed as
\[
\tilde\epsilon_{\rm eff} (\bar{\bf k})=g^{-1}\epsilon_{\rm eff} (\bar{\bf k})g.
\]
Using the above transformation rules, one can indeed
verify that our EOM (\ref{EOM1}, \ref{EOM2}) are invariant
under $SU(N)$ gauge transformation (\ref{GT}).

\sb{Abelian case : comparison with other approaches}
\ind
In the Abelian case : $N=1$, the above equations of motion
(\ref{EOM1},\ref{EOM2}) reduces to
\bea
{d\bar{x}_\mu\over dt} &=&
{\rd\epsilon_{\rm eff}(\bar{\bf k},\bar{\bf x},t)\over \rd\bar{k}_\mu}-
{\cal F}_{\bar{k}_\mu \bar{k}_\nu}{d\bar{k}_\nu\over dt}-
{\cal F}_{\bar{k}_\mu \bar{x}_\nu}{d\bar{x}_\nu\over dt}-
{\cal F}_{\bar{k}_\mu t}
\label{u1x} \\
{d\bar{k}_\mu\over dt} &=&-
{\rd\epsilon_{\rm loc}(\bar{\bf k},\bar{\bf x},t)\over\rd\bar{x}_\mu}
\label{u1k}
\eea
EOM similar to Eqs. (\ref{u1x},\ref{u1k}) have been derived,
to our knowledge, twice, using either
\ben
\item
Time-dependent variational principle \cite{SN} or,
\item
Path-integral method using Wannier basis.
\fn{We had some difficulty to jusitify
the use of Wannier basis used in Ref. \cite{KT}
as a {\it complete} basis necessary in the path
integral formalism.
This is closely related to the arbitrariness of Wannier
function discussed extensively in Ref. \cite{vdB1}}
\een
If we compare Eqs. (2.19) of Ref. \cite{SN} and our Eqs.
(\ref{u1x},\ref{u1k}), it can be observed that three terms,
\be
{\cal F}_{\bar{x}_\mu \bar{x}_\nu}{d\bar{x}_\nu\over dt}+
{\cal F}_{\bar{x}_\mu \bar{k}_\nu}{d\bar{k}_\nu\over dt}+
{\cal F}_{\bar{x}_\mu t}
\label{second}
\ee
are lacking on the right hand side of (\ref{u1k}).
However, one can easily check by a simple power counting
that these terms appear {\it only at orders higher than 2}
in the perturbation series w.r.t. $\beta$ or ${\bf x}-\bar{\bf x}$.
Let us briefly illustrate this point.
Since a subscript $\bar{x}$ implies a derivative w.r.t. $\bar{x}$,
which is always accompanied by ${\bf x}-\bar{\bf x}$, it increases
the power by one. 
It is also the case for the subscript $t$.
Therefore, the first and the last terms of (\ref{u1k})
turs out immediately to be at least of the second order
of $\beta$.
In the second line, i.e., in Eq. (\ref{u1k}),
since $\rd \epsilon_{\rm loc}
(\bar{\bf k},\bar{\bf x},t)/\rd\bar{x}_\mu$
is also of the first order w.r.t. $\beta$, 
one can verify that $d\bar{k}_\mu/dt$ is at least of the first order 
w.r.t. $\beta$, even if (\ref{second}) is added to it.
Taking everything into account, one can conclude
that the lacking terms (\ref{second}) are at least of the second order
w.r.t. $\beta$ or ${\bf x}-\bar{\bf x}$.

Another difference between Eqs. (\ref{u1x},\ref{u1k}) and
Eqs. (2.19) of Ref. \cite{SN} is that 
in our EOM for $d\bar{\bf k}/dt$, 
the derivative $\rd/\rd\bar{x}_\mu$ applies to
$\epsilon_{\rm loc}(\bar{\bf k},\bar{\bf x},t)$ and not to
$\epsilon_{\rm eff}(\bar{\bf k},\bar{\bf x},t)$ as in Ref. \cite{SN}.
In our formalism, as is clearly shown in Eq. (\ref{dkdt}),
there is no room for $\Delta\epsilon(\bar{\bf k},\bar{\bf x},t)$
to enter the expression (\ref{u1k}).
Nevertheless, repeating the same type argument, i.e.,
the power counting for $\Delta\epsilon(\bar{\bf k},\bar{\bf x},t)$,
one can confirm that the contribution from
$\Delta\epsilon(\bar{\bf k},\bar{\bf x},t)$ is not physically
relevant at the first order of $\beta$ or of ${\bf x}-\bar{\bf x}$.

We have not only developed a systematic perturbation
theory w.r.t. $\beta$ or ${\bf x}-\bar{\bf x}$, but also we make no
approximation apart from the assumptions stated in Sec. 3.
Our calculation must be, therefore, {\it exact} at the first order
of perturbation theory. 
Since possible discrepancies
start only at the second order in the perturbation
series, our result, Eq. (\ref{u1x},\ref{u1k}) is not inconsistent
\fn{The calculation done in Ref. \cite{SN} is also first order,
in particular, their Hamiltonian, Eqs. (2.1), (2.14) and (2.15),
is first order,
but they kept all the possible Berry phase contribution
without making further consideration of power counting,
whereas we omitted systematically higher order terms.}
with that of Ref. \cite{SN}.

\sa{Discussion : Berry phase engineering}

\ind
The gauge invariant EOM (\ref{EOM1},\ref{EOM2}) have been 
successfully derived in the previous section.
The decomposition (\ref{dec}) uncovered the orgin of two different 
types of reciprocal fields introduced in Sec. 2.
In this section we discuss some physical consequences of
Sec. 4 in the context of Berry phase engineering.

In the Introduction, we argued that a finite net
charge current could be induced by the $U(1)$ Berry phase correction
to the semiclassical EOM (\ref{C1},\ref{C2}). This finite charge current 
is actually carried by all the electrons below the Fermi surface,
i.e., by the electrons in the ground state.
Generalizing this $U(1)$ argument to the non-Abelian case,
we will discuss in this section how the various types
of non-abelian field strength appearing in our EOM are
related to concrete physical realizations,
mainly focusing on the $SU(2)$ case.
This opens a new possibility of manipulating the
ground state electronic wave function
by controlling Berry phase, which is sometimes called,
Berry phase engineering.

After introducing some terminologies and fixing notations,
we will focus on two topics.
In Sec. 5.2, we will see that ${\cal F}_{k_{\mu}k_{\nu}}$
is related to the physics of Hall type current.
We first observe that the charge Hall current can be described 
by a trace of non-abelian field strength 
${\cal F}_{k_{\mu}k_{\nu}}$, while this current vanishes 
whenever the system is time-reversally ($T$-) invariant.
The charge Hall current carried by a $({\bf k},\uparrow)$ Bloch 
electron and that of the $(-{\bf k},\downarrow)$ electron 
precisely cancel each other.
Based on this observation, we then propose
two physical situations in which the difficulty of the cancellation
of Hall type charge current will be overcome and the reciprocal
magnetic field manifests itself experimentally as a Hall type current.
The two situations are :
\ben
\item
anomalous (charge) Hall current observed in systems
with broken reversal symmetry, i.e., in ferromagnets,
\cite{CN,OMN,RS1,JNM}
\item
spin Hall current in time reversally symmetric systems.
\cite{hirsch,zhang,MNZ,sinova,culcer}
\een

In Sec. 5.3, we will argue that ${\cal F}_{k_{\mu}t}$ is 
directly related to various types of polarization currents, 
currents induced in insulators under time-dependent perturbations.     
Similarly to the case of Hall type charge current, which will be
discussed in Sec. 5.2, we first observe that in such systems
that are symmetric under spatial inversion,
the polarization electric/spin current actually vanishes
due to a cancellation associated with the inversion
symmetry of the system.
Then, in order to overcome this difficulty
we propose, in parallel with Sec. 5.2,
two physical systems in which the problem of cancellation
will be resolved, and the gauge invariant reciprocal
field strength appears explicitly in a macroscopic physcial
quantity, i.e., as a polarization current:
\ben
\item
If the inversion symmetry is broken externally or
spontaneously, the abelian ${\cal F}_{k_{\mu}t}$ gives rise
to a relevant contribution to the polarization electric/spin
current.
\item
Even in systems symmetric under spatial inversion,
non-abelian ${\cal F}_{k_{\mu}t}$ may have a chance
to manifest itself as a polarization {\it orbital} current,
if that orbital degree of freedom changes its sign under the 
spatial inversion.
\een
The analogy and correspondence between the Hall type
and polarization currents are summarized in Tables 1 and 2.

\sb{Preliminaries}
\ind
Before further discussing Berry phase transport, 
we first introduce some terminologies as well as giving an
unambiguous definition to spin/orbital currents. 

In order to illustrate our point, let us first consider a
spin current.
We have naturally in mind that there are different 
points of view \cite{MNZ,sinova,culcer}
on the definition of spin current operator,
$J^{\cal S}_{\alpha\mu}$.
The difficulty of defininig a spin current stems
simply from the fact that the bare spin $S_\alpha$ is generally 
{\it not} a conserved quantity due to spin-orbit interaction,
i.e., the continuity eqation,
$\rd S_\alpha/\rd t+
\sum_{\mu=1}^D \rd J^{\cal S}_{\mu\alpha}/\rd x_\mu=0$,
is not satisfied.
Thereby the Noether's theorem does not apply.
If one focuses on the time derivative of a local spin
$S_\alpha({\bf x},t)$ in the general case of non-conserved
spin, one could observe that there are two contributions
to it of physically different nature, i.e., contributions
from (i) a local spin curent, (ii) a local precession of spin.
The former is the one of our interest, and the
latter is related to the non-conservation of spin.
Unfortunately, there is no systematic prescription for distinguishing 
between those two contributions.

We can still define on quite general ground a current operator
${\bf J}_{\cal I}$ 
associated with an internal degree of freedom ${\cal I}$
as the time derivative of a spatial polarization of ${\cal I}$ 
as,
\bea 
{\bf J}_{\cal I} &=& \frac{d{\bf P}_{\cal I}}{dt},
\label{JS1} \\  
{\bf P}_{\cal I}&=&\frac{1}{L^{D}}\sum_{j=1}^{M}\ 
\frac{1}{2}\Big\{{\cal I}_j
{\bf x}_j+ {\bf x}_j{\cal I}_j\Big\}.
\label{JS2}
\eea
$L$ and $M$ denote the system size and the total number of electrons, 
respectively. 
The subscripts $j$ attributed to ${\cal I}$ and ${\bf x}$ specify
each electron. 
In the case of a spin current, the operator ${\cal I}$ should
be replaced by a usual spin operator, ${\cal I}=S_\alpha$.
The main reason why we have defined the spin/orbital current as
Eq. (\ref{JS1},\ref{JS2}) is that that is the one which could be
diretcly observable.
\fn{Otherwise, we could have defined it also as
\[
\tilde{\bf J}_{\cal I}=\frac{1}{L^{D}}\sum_{j=1}^{M}\ 
\frac{1}{2}\bigg\{
{\cal I}_j{d{\bf x}\over dt}+{d{\bf x}\over dt}{\cal I}_j
\bigg\}. 
\] This type of definition is convenient for the application of
Kubo formula. \cite{culcer,sinova,loss}} 
For example, an increasing 
$\big(P_{S_\alpha}\big)_\mu=P_\mu(S_\alpha)$ indicates that    
extra up (down)-spin electrons with $S_\alpha=+1/2$ $(-1/2)$
acculmulate in one (the other) end of a system with 
$x_\mu=L/2$ $(-L/2)$, which could be 
experimentally detected by some optical probes. 

We will also discuss orbital currents. 
A Bloch electron has, in addition to the spin degree of freedom,
orbital degrees of freedom $\Pi$ in multiband systems. 
These orbital degrees of freedom describe the charge distribution 
in the unit cell, and hence,
\fn{The periodicity of $\Pi$, i.e., Eq. (\ref{pia}) 
excludes its off-diagonal matrix elements in ${\bf k}$-space:   
\bea
&&\la\Pi\ra_{mn}({\bf k}^{\prime},{\bf k}) 
= \delta({\bf k}^{\prime}-{\bf k}) \nn  
\la\Pi\ra_{mn}({\bf k}) \nn \\ 
&&\la\Pi\ra_{mn}({\bf k}) 
= \Bla\phi_{m}({\bf k},t)
\Big|\Pi \Big|\phi_{n}({\bf k},t)
\Bra \nn \\ 
&&=\frac{(2\pi )^{D}}{V_{\rm cell}} 
\int_{\rm unit\ cell}d{\bf x}\  \Big
\la u_{m}({\bf k},t)\Big|{\bf x}\Bra 
\Pi\Big({\bf x},-i\nabla +{\bf k}\Big)
\Bla{\bf x}\Big|u_{n}({\bf k},t)\Bra.
\label{pimn}
\eea}
\be
\Pi ({\bf x},{\bf p})=\Pi ({\bf x}+{\bf a},{\bf p}). 
\label{pia}
\ee
In Sec. 5.3 we focus on an orbital operator $\Pi$
which behaves quite contrastingly to the spin under
time reversal and spatial inversion, i.e.,
\bea
\Pi ({\bf x},{\bf p})&=&-\Pi (-{\bf x},-{\bf p}), 
\label{Ipi} \\
\Pi ({\bf x},{\bf p})&=&\Pi ^{\ast}({\bf x},{\bf p}). 
\label{pi*}
\eea
In contrast to the spin operator $S$,
the parity operator $\Pi$
reverses its sign under spatial inversion, while 
invariant under time reversal.
Accordingly, we dub this oribtal operator as 
a {\it parity} operator.
\fn{This orbital operator is different from the angular momentum 
operator, ${\bf\omega}={\bf x}\times{\bf p}$, which we might also call 
an orbital operator.
${\bf\omega}$ reverses its sign under 
time reversal, while remains invariant under spatial inversion.}
Various transformation properties of the operators ${\cal I}=S_\alpha,\Pi $
are summarized in Table 1.

The spin/orbital current associated with an operator ${\cal I}$
was introduced in Eqs. (\ref{JS1},\ref{JS2}). 
We calculate in Secs. 5.2 and 5.3 those spin/parity currents 
carried by the ground state. 
In order to make later discussions clearer and physically more 
appealing, we make the following assumptions:
\ben
\item
We keep only their matrix elements in the restricted subspace
${\cal N}$ spanned by $N$-fold degenerate bands ;
If $m\in {\cal N}$ and $l\not\in {\cal N}$, then
\bea
\Bla\phi_{m}({\bf k})\Big|S_\alpha
\Big|\phi_{l}({\bf k})\Bra &=& 0, 
\label{as1} \\
\Bla\phi_{m}({\bf k})\Big|\Pi ({\bf x},{\bf p})
\Big|\phi_{l}({\bf k})\Bra &=& 0. 
\label{as2}
\eea
\een
As will be seen in Secs. 5.2 and 5.3, 
these approximations make the following discussions
considerably simpler, i.e.,
not only a charge current 
but also spin/parity currents become related simply
to the non-abelian field strength, 
${\cal F}_{k_{\mu}k_{\nu}}$ and ${\cal F}_{k_{\mu}t}$.

\sb{${\cal F}_{k_{\mu}k_{\nu}}$ induces Hall type currents:
AHE and spin Hall effect}

\ind
The field strength ${\cal F}_{k_{\mu}k_{\nu}}$ describes 
various kinds of spontaneous Hall currents carried by 
the ground state. 
In order to demonstrate it we expose our system
under a uniform electric field ${\bf E}$;
we consider below the following Hamiltonian,
\bea
H({\bf p},{\bf x},\beta({\bf x},t))
= H_{0}({\bf p},{\bf x})+e\sum_{\mu=1}^{D} E_\mu x_\mu. 
\label{ham52}
\eea
We see below that 
the Hall-type current can be expressed essentially as
a trace of ${\cal I}{\cal F}_{k_{\mu}k_{\nu}}$
in the $N$-fold degenerate pseudospin space.
The cancellation or the survival
of such Hall-type topological current is essentially related to 
the transformation properties under time reversal ($T$)
of the system.
In systems with broken $T$ symmetry
a finite charge/mass Hall current generally appear.
In two spatial dimension $D=2$, this situation is often
described in terms of Chern-Simons gauge field,
accounting for the quantized Hall conductance,
\cite{kenzo}
as well as fractional charge and statistics.
\cite{SCZ}
Chern-Simons terms also appear in electrically neutral systems.
\cite{goryo}
Quantization of spin Hall conductance in unconventional superfluids 
has also been studied in this context.
\cite{senthil}

\sc{Charge Hall current}

\ind
Let us first see that the trace of the non-abelian 
field strength ${\cal F}_{k_{\mu}k_{\nu}}$ 
describes a spontaneous charge Hall conductivity, 
i.e., an anomalous Hall conductivity.  
\cite{CN,OMN,RS1,JNM}
In terms of Eqs. (\ref{JS1},\ref{JS2}), we are considering 
the case of ${\cal I}={\bf 1}$.
Applying the EOM, Eqs. (\ref{EOM1},\ref{EOM2}), 
to the present case, we consider
a Bloch electron, in the $j$-th pseudospin state 
\fn{The $j$-th eiegenstate is a linear combination of
different pseudospin states $m=1,\cdots,N$, i.e., 
$\sum_{m=1}^{N}z^{(j)}_{m}\ |\phi_{m}(\bar{\bf k}) \ra$. 
These eigenstates are chosen to be orthogonal to each other:  
${\bf z}^{(i)\dagger}{\bf z}^{(j)} = \delta_{ij}$ for
$i,j=1,\cdots,N$.}
and with a crystal momentum $\bar{\bf k}$. 
The $SU(N)$ gauge invariant EOM for this electron
under a uniform electric field ${\bf E}$ now reads,
\bea
\frac{d\bar{x}^{(j)}_{\mu}}{dt}&=&
\frac{\rd \epsilon_{\rm loc}(\bar{\bf k})}{\rd \bar{k}_{\mu}} - 
\sum _{\nu=1}^{D}{\bf z}^{(j)\dagger}   
{\cal F}_{{k}_{\mu}{k}_{\nu}}(\bar{\bf k})   
{\bf z}^{(j)}\ \frac{d \bar{k}_{\nu}}{dt},
\label{EOM52x} \\ 
\frac{d \bar{k}_{\mu}}{dt} &=& -eE_\mu. 
\label{EOM52k}
\eea 
When the crystal momentum $\bar{\bf k}$ is located below the Fermi 
surface, all those pseudospin states are completely 
occupied and contribute to the charge current.
In order to calculate the charge Hall current, we need
\bea
-e\sum_{j=1}^{N}\frac{d\bar{x}^{(j)}_{\mu}}{dt} 
&=&-N e \frac{\rd \epsilon_{\rm loc}(\bar{\bf k})}{\rd \bar{k}_{\mu}}-e^2
\sum_{j=1}^{N}{\bf z}^{(j)\dagger}
{\cal F}_{k_{\mu}k_{\nu}}(\bar{\bf k}){\bf z}^{(j)}E_{\nu} 
\nn \\
&=& - N e \frac{\rd \epsilon_{\rm loc}(\bar{\bf k})}{\rd \bar{k}_{\mu}} - e^{2}\ 
{\rm  Tr}\ \Big[{\cal F}_{k_{\mu}k_{\nu}}(\bar{\bf k})\Big]\ E_{\nu}.
\label{CH1} 
\eea 
Then, by integrating these contributions over filled $\bar{\bf k}$ 
points, we obtain the total current carried by all the electrons 
below the Fermi energy $\epsilon_{\rm F}$,  
\be
\big( J^{\rm Hall}_{\cal C}\big)_{\mu}=  
-e^{2}\ \sum_{\nu=1}^{D}
\int_{\epsilon_{\rm loc}(\bar{\bf k})<\epsilon_{\rm F}} 
\frac{d\bar{\bf k}}{(2\pi )^{D}}\ 
{\rm Tr}\ \Big[{\cal F}_{k_{\mu}k_{\nu}}(\bar{\bf k})\Big]\ 
E_{\nu}.  
\label{JCH} 
\ee
Here we assumed for the sake of simplicity that 
the non-perturbed Hamiltonian $H_{0}({\bf p},{\bf x})$ 
is also invariant under spatial inversion ($I$).
As far as the Hall type current is concerned, however,
this $I$ symmetry plays only a minor role.
Eq. (\ref{JCH}) takes indeed the form of a Hall type current 
reflecting the antisymmetry of 
${\cal F}_{k_{\mu}k_{\nu}}$: 
${\cal F}_{k_{\mu}k_{\nu}}=-{\cal F}_{k_{\nu}k_{\mu}}$.  
The first term of Eq. (\ref{CH1}) did not 
contribute to Eq. (\ref{JCH}) due to a cancellation 
associated with the $I$ symmetry. 
On the contrary, Eq. (\ref{JCH}) 
remains finite irrespective of the $I$-invariance of $H_{0}$.  

Unfortunately, the charge Hall current obtained in Eq.
(\ref{JCH}) vanishes whenever the system is $T$-invariant. 
Let us see this point more explicitly. 
We consider an unperturbed Hamiltonian
$H_{0}({\bf x},{\bf p})$ which is {\it invariant} under $T$.  
The $T$-invariance relates its 
$N$-fold degenerate Bloch functions at ${\bf k}$ with those at 
$-{\bf k}$ upto a certain $SU(N)$ gauge transformations $g({\bf k})$,
\bea
\sum_{b=\uparrow,\downarrow}
\big[i\sigma_{y}\big]_{ab}
\bla {\bf x},b \big|u_{i}
(-{\bf k})\bra^{\ast} = \sum_{j=1}^{N}
\bla {\bf x},a \big|u_{j}({\bf k})
\bra g_{ji}({\bf k}). 
\label{g}
\eea
In the language of field strength, this reduces to, 
\be
{\cal F}_{k_{\mu}k_{\nu}}(-{\bf k})^{t}=-
g^{-1}{\cal F}_{k_{\mu}k_{\nu}}({\bf k}) g,  
\label{gFkk} 
\ee
where the superscript $t$ represents a transposed matrix,
i.e., $\big({\cal F}^t\big)_{mn}=({\cal F})_{nm}$.
One can verify this using Eqs. (\ref{vec},\ref{field},\ref{g}). 
Consequently, the charge current carried 
by a Bloch electron at ${\bf k}$ cancels with that  
of $-{\bf k}$ electron, i.e., 
\bea 
\sum_{\nu=1}^{D}{\rm Tr} \Big[{\cal F}_{k_{\mu}k_{\nu}}({\bf k})\Big] 
E_{\nu}= - \sum_{\nu=1}^{D} 
{\rm Tr} \Big[{\cal F}_{k_{\mu}k_{\nu}}(-{\bf k})\Big]
E_{\nu}.
\label{trF} 
\eea
Eqs. (\ref{JCH},\ref{trF}) indicate the absence of spontaneous Hall current 
in $T$-invariant systems.

\sc{Spin Hall current}

\ind
We are thus led to investigate the {\it spin} current, expecting
that the spin Hall current remains finite even in $T$-invariant systems.
\cite{hirsch,zhang,MNZ,sinova,culcer}
The underlying idea is simply that 
such a sign change as seen in Eq. (\ref{trF}) may be compensated
by that of spin operator under the operation of time reversal.
The spin current has been defined in Eq. (\ref{JS1}), 
in a more general context for an arbitrary internal degree of 
freedom ${\cal I}$.
Let us first observe that the $T$-invariance of 
$H_{0}({\bf x},{\bf p})$ is instrumental for this compensation.
The total spin carried by the Bloch electrons at ${\bf k}$ 
has the same absolute value and {\it opposite sign} 
of that of the Bloch electrons at $-{\bf k}$,
\fn{To be more specific, $T$-invariance relates the
matrix of $S_\alpha$ at ${\bf k}$ and at $-{\bf k}$ 
through a $SU(N)$ gauge transformation $g$, 
\bea
\l\la S_\alpha\r\ra(-{\bf k})^{t}=-g^{-1}  
\l\la S_\alpha\r\ra({\bf k})g. 
\label{gS}
\eea
Eq. (\ref{gS}) can be verified explicitly using Eq. (\ref{g}).}
\be
{\rm Tr}\big[\l\la S_\alpha\r\ra ({\bf k})\big]=- 
{\rm Tr}\big[\l\la S_\alpha\r\ra (-{\bf k})\big],
\label{trS}
\ee
where $\l\la S_\alpha\r\ra({\bf k})$ is a $N\times N$ matrix,
whose $(m,n)$-components are given by
\bea
&&\bla S_\alpha\bra_{mn}({\bf k})=
\bla\phi_{m}({\bf k})\big|S_\alpha\big|\phi_{n}({\bf k})\bra
\nn \\
&=&\frac{(2\pi )^{D}}{V_{\rm cell}} 
\int_{\rm unit\ cell}d{\bf x}
\sum_{a,b=\uparrow,\downarrow}\bla u_{m}({\bf k})\big|{\bf x},a\bra 
\frac{1}{2}\big(\sigma_\alpha\big)_{ab}
\bla {\bf x},b \big| u_{n}({\bf k})\bra.
\label{Smn}
\eea
Let us now consider the spin Hall current defined as 
(\ref{JS1},\ref{JS2}) with ${\cal I}$ being ${\cal I}=S_\alpha$,
the usual spin operator.
In order to evaluate a {\it spin} Hall current, 
the EOM (\ref{EOM52x},\ref{EOM52k}) used for the calculation of 
{\it charge} Hall current are no longer sufficient.
We need instead to derive EOM for an observable, 
\be
{\cal O}_{\mu\alpha}^{\cal S}=
{\cal O}_\mu(S_\alpha)=
\frac{1}{2}
\big(S_\alpha x_\mu+x_\mu S_\alpha\big).   
\label{OS}
\ee 
Following the same type of procedure as 
the derivation of Eqs. (\ref{EOM52x},\ref{EOM52k}),
we perform, in particular, the decomposition (\ref{dec}). 
As was also the case in Eqs. (\ref{EOM52x},\ref{EOM52k}),
the second contribution to (\ref{dec}), i.e.,
$\Omega^{(2)}_{\cal O}$ vanishes 
in the present case.
\fn{As long as the electric field ${\bf E}$ in Eq. (\ref{ham52})
is uniform, the local Bloch function defined in 
Eq. (\ref{Hloc}) has no dependence on $\bar{\bf x}$ and $t$. 
As a result, $\Omega^{(2)}_{\cal O}$ vanishes.}
The EOM reads,
\bea
&&\frac{d\bar{\cal O}_{\mu\alpha}^{\cal S}}{dt} 
= i\sum_{m,n=1}^{N}\int d{\bf k}d{\bf k}^{\prime}
a^{\ast}_{m}({\bf k},t)
\Big[\bla H\bra ,\bla {\cal O}_{\mu\alpha}^{\cal S}\bra\Big]_{mn}
({\bf k},{\bf k}^{\prime})
a_{n}({\bf k}^{\prime},t)
\nn \\
&&=i\sum_{m,n=1}^{N}\int 
d{\bf k}a^{\ast}_{m}
\Big[\bla H_{0}\bra +
e\sum_{\mu=1}^{D}E_\mu\bla x_\mu \bra,
\l\la {\cal O}_{\mu\alpha}^{\cal S}\r\ra \Big]_{mn}a_{n}.
\label{SH1} 
\eea
The assumption (\ref{as1}) allows us to rewrite Eq. (\ref{SH1}) as
\bea
&&\frac{d\bar{\cal O}_{\mu\alpha}^{\cal S}}{dt}= 
\sum_{m,n=1}^{N} \int d{\bf k} 
\rho({\bf k}) z^{\ast}_{m}({\bf k},t)\Big[
\l\la S_\alpha\r\ra_{mn}({\bf k})\ 
\frac{\rd \epsilon_{\rm loc}(\bar{\bf k})}{\rd k_{\nu}} 
\nn \\ 
&&+\frac{e}{2}\sum_{\nu=1}^{D}\Big\{
\Big(\l\la S_\alpha\r\ra({\bf k}) 
{\cal F}_{k_\mu k_\nu}({\bf k})\Big)_{mn}
E_{\nu}+{\rm h.c.}\Big\}\Big]z_{n}({\bf k},t).
\label{SH2}
\eea
The details of the derivation of Eq. (\ref{SH2}) is given in
Appendix C.
We then apply the prescription (\ref{presc}), 
replacing ${\bf k}$ in the integrant by its mean value 
$\bar{\bf k}$. 
The contribution to the spin Hall current by an electron 
occupying the $j$-th pseudospin state at $\bar{\bf k}$
is, therfore,
\bea
\frac{d\bar{\cal O}_{\mu\alpha}^{{\cal S}(j)}}{dt}&=& 
{\bf z}^{(j)\dagger}
\l\la S_\alpha\r\ra(\bar{\bf k}) {\bf z}^{(j)}\  
\frac{\rd \epsilon_{\rm loc}(\bar{\bf k})}{\rd \bar{k}_\mu}
\nn \\ 
&+&\frac{e}{2}\sum_{\nu=1}^{D}
\bigg\{{\bf z}^{(j)\dagger}
\la S_\alpha\ra(\bar{\bf k})
{\cal F}_{k_\mu k_\nu}(\bar{\bf k}) 
{\bf z}^{(j)} + {\rm c.c.}
\bigg\}\ E_{\nu}.
\label{SH3} 
\eea
Finally we take the summation over $N$ pseudospin states and 
over filled $\bar{\bf k}$ points to find,
\bea
\big(J_{\cal S}^{\rm Hall}\big)_{\mu\alpha}&=& 
\int_{\epsilon_{\rm loc}(\bar{\bf k})<\epsilon_{\rm F}} 
\frac{d\bar{\bf k}}{(2\pi )^{D}} 
\sum_{j=1}^{N} {d\bar{\cal O}_{\mu\alpha}^{{\cal S}(j)}\over dt} 
\nn \\
&=&e\ \sum_{\nu=1}^{D}\ 
\int_{\epsilon_{\rm loc}(\bar{\bf k})<\epsilon_{\rm F}} 
\frac{d\bar{\bf k}}{(2\pi )^{D}}  
\ {\rm Tr}\Big[ \l\la S_\alpha\r\ra (\bar{\bf k})
{\cal F}_{k_\mu k_\nu}(\bar{\bf k}) \Big]
\ E_{\nu}. 
\label{JSH} 
\eea
As was the case in Eq. (\ref{JCH}), we also assumed in
Eq. (\ref{JSH}) that $H_0$ is invariant under $I$,
so that the first term of Eq. (\ref{SH3}) should not
appear in Eq. (\ref{JSH}), i.e.,
\be
\int_{\epsilon_{\rm loc}(\bar{\bf k})<\epsilon_{\rm F}} 
\frac{d\bar{\bf k}}{(2\pi )^{D}}
{\rm Tr}\Big[\l\la S_\alpha\r\ra({\bf k})\Big]
\frac{\rd \epsilon_{\rm loc}(\bar{\bf k})}{\rd \bar{k}_{\nu}} = 0.
\label{JSH1}
\ee
Contrary to this {\it normal} part, the spin Hall current 
associated with the anomalous velocity, given by Eq. (\ref{JSH}), 
has a possibility to be finite irrespective of the $I$ and $T$ 
symmetries.
When the system is invariant under either $T$ or $I$,
\fn{In the $I$-invariant case, instead of Eq. (\ref{g}), 
Eq. (\ref{h}) holds, i.e., the $I$-invariance of $H_0$ relates
$\bla -{\bf x},a \big|u_{i}(-{\bf k})\bra$
with $\bla {\bf x},\tau \big|u_{j}({\bf k})\bra$
up to a $SU(N)$ gauge degree of freedom $h$.   
This reduces in terms of spin and field strength to
\bea
\l\la S_\alpha\r\ra(-{\bf k})&=&
h^{-1} \l\la S_\alpha\r\ra({\bf k}) h,
\label{hS} \\  
{\cal F}_{k_{\mu}k_{\nu}}(-{\bf k})&=&
h^{-1} {\cal F}_{k_{\mu}k_{\nu}}({\bf k}) h. 
\label{hFkk}
\eea
Eq. (\ref{hS}) justifies (\ref{JSH1}), whereas multiplying 
Eqs. (\ref{hS}) and (\ref{hFkk}), one finds immediately
Eq. (\ref{trSF}).}
the spin current carried by a Bloch electron at ${\bf k}$ and
that of $-{\bf k}$ give the same contribution;
\bea
{\rm Tr}\Big[\l\la S_\alpha\r\ra({\bf k})
{\cal F}_{k_\mu k_\nu}({\bf k})\Big] 
= {\rm Tr}\Big[\l\la S_\alpha\r\ra(-{\bf k}) 
{\cal F}_{k_\mu k_\nu}(-{\bf k})\Big].
\label{trSF} 
\eea
Under $T$ symmetry Eq. (\ref{trSF}) is a consequence of
Eqs. (\ref{gFkk},\ref{gS}).
Eqs. (\ref{JSH},\ref{trSF}) confirm that our expectation that the
spin Hall current is robust against $T$ symmetry was indeed the
case.
A set of cancellation rules for the Hall type currents are
established in Table 2.

\sb{${\cal F}_{k_{\mu}t}$ induces a polarization current:
parity polarization current and quantum spin pump}

\ind
We have seen in the previous section 
that ${\cal F}_{k_{\mu}k_{\nu}}$ is related to Hall type
currents associated with the internal degrees of 
freedom such as charge and spin,
by applying a uniform electric field to the system. 
Here we argue that 
${\cal F}_{k_{\mu}t}$ describes various kinds of 
{\it polarization} current. 
More specifically, we consider a situation where a band 
insulator is subject to a time-dependent 
perturbation which {\it does not} break the periodicity of 
the underlying crystal, i.e., we consider a Hamiltonian,
\be
H({\bf x},{\bf p};\beta(t))= 
H({\bf x}+{\bf a},{\bf p};\beta(t)).
\label{ham53}
\ee
Since the perturbation $\beta(t)$ does not depend on ${\bf x}$, the {\it local}
Hamiltonian defined in Eq. (\ref{Hloc}) reduces simply to
\[
H_{\rm loc}({\bf x},{\bf p},t)=H({\bf x},{\bf p};\beta(t)).
\]
Depending on the perturbation $\beta(t)$, 
the ground state wave function of the local Hamiltonian, 
$H_{\rm loc}({\bf x},{\bf p},t)$ 
also evolves temporally.   
Since an electronic wave function for the ground state describes 
spatial distributions of charge, 
spin and orbital, its evolution in general induces 
various kinds of currents in the system.  
When the system is isolated from the external 
circuit, an induced current accumulates an extra charge
(or spin, orbital) on one side of the system,
which results in a spatial polarization of charge, spin and 
orbital. \cite{vdB2} 
Accordingly, this type of current associated with such internal 
degrees of freedom is often called a {\it polarization} current.
In the following, we describe the physics of polarization current 
using the language of non-abelian gauge field, in particular, 
that of ${\cal F}_{k_{\mu}t}$.
One of the advantges of taking such a viewpoint is that the role
of symmetry becomes transparent, which we summarized as a set of
{\it cancellation rules} in Table 2.

\sc{Charge polarization current}

\ind
Let us first consider a {\it charge} polarization current.
In the case of time-dependent perturbation (\ref{ham53}),
the EOM analogous to Eqs. (\ref{EOM52x},\ref{EOM52k}), 
are found to be  
\bea
\frac{d\bar{x}^{(j)}_{\mu}}{dt} &=& \frac{\rd \epsilon_{\rm loc}({\bf k})} 
{\rd \bar{k}_{\mu}} - {\bf z}^{(j)\dagger} 
 {\cal F}_{k_{\mu}t}(\bar{\bf k},t)  {\bf z}^{(j)}, 
\label{EOM53x} \\
\frac{d\bar{k}^{(j)}_{\mu}}{dt} &=& 0,
\label{EOM53k}  
\eea
where $j = 1,\cdots,N$.
Collecting contributions from all $N$ pseudospin states and 
from all filled $\bar{\bf k}$ points, one can calculate
the charge current carried by the ground state as
\bea 
\big({\bf J}_{\cal C}^{\rm pol}\big)_{\mu} &=& 
-e \int_{\rm BZ} \frac{d\bar{\bf k}}{(2\pi )^{D}}
\sum_{j=1}^{N}\frac{d \bar{x}^{(j)}_{\mu}}{dt} 
\nn \\
&=& e\int_{\rm BZ} \frac{d\bar{\bf k}}{(2\pi )^{D}}\  
{\rm Tr} \Big[{\cal F}_{k_{\mu}t}
(\bar{\bf k},t)\Big],
\label{JCP} 
\eea 
where the $\bar{\bf k}$-integral was performed over the whole  
Brillouin zone (BZ).
Eq. (\ref{JCP}) is analogous to Eq. (\ref{JCH}), which we found
for the charge Hall current.
We can see that the trace of different types of reciprocal field 
strength, i.e., ${\cal F}_{k_\mu t}$ and ${\cal F}_{k_\mu k_\nu}$
are related to different types of physical currents,
i.e., polarization and Hall type currents.

We have seen in the previous section that
no charge {\it Hall} current flows whenever the sysytem is invariant
under time reversal $T$.
In contrast, we are going to see below that
the charge {\it polarization} current vanishes
whenever the system is invariant under spatial inversion $I$,
\fn{i.e., the underlying crystal structure has centro-symmetric 
points.}
\bea
H_{\rm loc}({\bf x},{\bf p},t)=H_{\rm loc}(-{\bf x},-{\bf p},t).
\eea
In this case, its $N$-fold degenerate Bloch functions 
at ${\bf k}$ is related to those at $-{\bf k}$ upto a certain 
$SU(N)$ gauge transformation $h({\bf k},t)$,  
\bea
\bla -{\bf x},a \big|u_{i}(-{\bf k},t)\bra  
= \sum_{j=1}^{N}
\bla {\bf x},a \big|u_{j}({\bf k},t)\bra
h_{ji}({\bf k},t). 
\label{h}
\eea 
Since the field strength ${\cal F}_{k_{\mu}t}$ 
is related through Eqs. (\ref{vec},\ref{field}) to those wavefunctions,
Eq. (\ref{h}) reduces to the following identity,  
\bea
{\cal F}_{k_{\mu}t}(-{\bf k},t) =
 - h^{-1} \   
{\cal F}_{k_{\mu}t}({\bf k},t) h. 
\label{hFkt} 
\eea
Consequently,  
\be
{\rm Tr} \Big[{\cal F}_{k_{\mu}t}(-{\bf k},t)\Big]=- 
{\rm Tr} \Big[{\cal F}_{k_{\mu}t}({\bf k},t)\Big].  
\label{trFkt} 
\ee
Eqs. (\ref{JCP}) and (\ref{trFkt}) indicate that 
the {\it charge} polarization current always vanishes in
$I$-invariant systems.

\sc{Parity polarization current}

\ind
We have already encountered a similar 
situation in the previous section. Under the time reversal $T$,
${\cal F}_{k_{\mu}k_{\nu}}({\bf k})$ is transformed to 
$-\big({\cal F}_{k_{\mu}k_{\nu}}(-{\bf k})\big)^{t}$ 
upto a $SU(N)$ gauge degree of 
freedom, as can seen in Eq. (\ref{gFkk}). 
As a result, the {\it charge} Hall current 
vanished in $T$-invariant systems. On the other hand, 
a {\it spin} Hall current was robust against $T$-invariance. 
The reason was that
not only ${\cal F}_{k_{\mu}k_{\nu}}$ but also 
the spin operator are odd under the time-reversal, as
given respectively in Eqs. (\ref{gFkk}) and (\ref{gS}). 

Following the same type of logic, we can expect that 
an {\it orbital} polarization current may remain finite 
irrespective of the $I$-invariance of $H_{\rm loc}$, 
as far as the associated orbital operator 
$\Pi({\bf x},{\bf p})$
changes its sign under the spatial inversion $I$.
\fn{See Eq. (\ref{Ipi}).}
Accordingly we dub this type of orbital current a
{\it parity} polarization current.   

Expecting that the above analogy is indeed a sensible one, 
let us further analyze the parity polarization current carried 
by the ground state. 
Since we have defined this orbital current as 
Eq. (\ref{JS1},\ref{JS2}), 
we have to consider an EOM for
\be
{\cal O}_\mu^\Pi =
{\cal O}_\mu\big(\Pi({\bf x},{\bf p})\big)=
{1\over 2}\big(\Pi x_\mu+x_\mu\Pi  \big).
\ee 
We derive their EOM in terms of the decomposition (\ref{dec}). 
Our local Hamiltonian is  time-dependent, 
and so is its local Bloch function. 
Therefore, $\Omega^{(2)}_{{\cal O}_\mu^\Pi }$ appearing 
in Eq. (\ref{dec}) remains finite in general: 
\bea
&&\frac{d\bar{\cal O}_\mu^\Pi }{dt} 
=\Omega^{(1)}_{{\cal O}_\mu^\Pi }+ 
\Omega^{(2)}_{{\cal O}_\mu^\Pi } 
\nn \\
&=&i\sum_{m,n=1}^{N}\int d{\bf k} d{\bf k}^{\prime} 
a^{\ast}_{m}({\bf k},t)
\Big[\l\la H_{\rm loc}\r\ra 
-i\nabla_{t},\l\la {\cal O}_\mu^\Pi \r\ra
\Big]_{mn}({\bf k},{\bf k}^{\prime})
a_{n}({\bf k}^{\prime},t). 
\label{PP1} 
\eea
This equation is analogous to Eq. (\ref{SH1}).
The covariant derivative 
$\nabla_{k_{\mu}}=\l\la x_\mu \r\ra$
in Eq. (\ref{SH1})
was replaced in Eq. (\ref{PP1}) by another covariant 
derivative w.r.t. time, i.e., $\nabla_{t}$:
\bea
\Big(i\nabla_{t}\Big)_{mn}=
\frac{\rd}{\rd t}\delta_{mn}-
i\Big({\cal A}_{t}\Big)_{mn}. 
\eea  
Let us further develop the analogy between the two cases,
i.e., we rewrite Eq. (\ref{PP1}) in the following way,
precisely as we rewrote (\ref{SH1}) as (\ref{SH2}).
The details of the derivation is given in Appendix D,
which is in parallel with Appendix C, and the result is
\bea
\frac{d\bar{\cal O}_\mu^\Pi }{dt}&=&
\sum_{m,n=1}^{N} \int d{\bf k}\rho({\bf k})
z^{\ast}_{m}({\bf k},t)
\bla \Pi \bra_{mn}({\bf k})
\frac{\rd \epsilon_{\rm loc}({\bf k})}{\rd k_{\mu}}
z_{n}({\bf k},t)
\nn \\
&-&{1\over 2}
z^{\ast}_{m}({\bf k},t)
\Big(\bla \Pi\bra({\bf k}){\cal F}_{k_{\mu}t}({\bf k})  
+{\rm h.c.}\Big)_{mn}
z_{n}({\bf k},t).
\label{PP2} 
\eea
Following the prescription (\ref{presc}), 
we see that the parity polarization current 
carried by $\bar{\bf k}$ Bloch electron occupying the 
$j$-th pseudospin state is given by, 
\bea
{d{\cal O}_\mu^{\Pi (j)}\over dt} 
&=&{\bf z}^{(j)\dagger}
\la \Pi \ra(\bar{\bf k}) {\bf z}^{(j)}\  
\frac{\rd \epsilon_{\rm loc}({\bar{\bf k}})}
{\rd \bar{k}_{\mu}}
\nn \\
&-&\Big({\bf z}^{(j)\dagger} 
\la \Pi \ra (\bar{\bf k}) 
{\cal F}_{k_{\mu}t}(\bar{\bf k}) 
{\bf z}^{(j)} + {\rm c.c.}\Big). 
\label{PP3}
\eea 
After taking its summation 
over $N$ pseudo-spin states and  over filled  
$\bar{\bf k}$ points, we finally obtain  
a parity polarization current carried by 
the ground state,  
\bea
\big({\bf J}_{\cal O}^{\rm pol}\big)_\mu&=&
\int_{\rm BZ}\frac{d\bar{\bf k}}{(2\pi )^{D}}
\sum_{j=1}^{N}
\ \frac{d\bar{\cal O}_\mu^{\Pi (j)}}{dt} \nn \\
&=& - \int_{\rm BZ}\frac{d\bar{\bf k}}{(2\pi )^{D}}\  
{\rm Tr}\Big[\ \la \Pi \ra (\bar{\bf k}) 
{\cal F}_{k_{\mu}t}(\bar{\bf k})\ \Big]. 
\label{JPP}
\eea  
The first term of Eq. (\ref{PP3}) did not contribute 
to Eq. (\ref{JPP}) due to a cancellation 
associated with the $T$-invariance,
\fn{$T$-invariance relates 
$N$-fold degenerate Bloch functions at ${\bf k}$ 
with the ones at $-{\bf k}$ upto a $SU(N)$ gauge degree of 
freedom $g$ as Eq. (\ref{g}).
This implies,
\bea
\Big(\la \Pi \ra(-{\bf k})\Big)^{t}
&=&g^{-1}\la \Pi \ra({\bf k})g, 
\label{gpi}
\\
\Big({\cal F}_{k_{\mu}t}(-{\bf k})\Big)^{t}
&=&g^{-1}{\cal F}_{k_{\mu}t}({\bf k}) g 
\label{gFkt}.
\eea
Eq. (\ref{gpi}) justifies (\ref{PP4}), whereas
multiplying Eq. (\ref{gpi}) with Eq. (\ref{gFkt})
one finds immediately (\ref{trpiF}).}
\be
\int_{\rm BZ} 
\frac{d\bar{\bf k}}{(2\pi )^{D}}\ {\rm Tr}
\Big[\la \Pi \ra(\bar{\bf k})\Big]\ 
\frac{\rd \epsilon_{\rm loc}(\bar{\bf k})}{\rd \bar{k}_{\mu}} = 0. 
\label{PP4}
\ee
On the contrary, the Berry phase contribution, i.e.,
Eq. (\ref{JPP}) turns to be quite robust against
both $I$ and $T$ symmetries.
In particular, in the $I$-invariant case, 
$\la \Pi \ra({\bf k})$ 
is identical to $-\la \Pi \ra (-{\bf k})$ up to 
the $SU(N)$ gauge transformation $h$,
\be
\la \Pi \ra(-{\bf k}) = - h^{-1}
\la \Pi \ra({\bf k})h  
\label{hpi}
\ee
This can be shown explicitly using Eqs. 
(\ref{Ipi},\ref{pi*},\ref{h},\ref{pimn}). 
Eqs. (\ref{hFkt},\ref{hpi}) indicate
\bea
{\rm Tr}\Big[\la \Pi \ra ({\bf k}) 
{\cal F}_{k_{\mu}t}({\bf k})\ \Big] 
= {\rm Tr}\Big[\ \la \Pi \ra (-{\bf k}) 
{\cal F}_{k_{\mu}t}(-{\bf k})\ \Big]. 
\label{trpiF}
\eea
Eq. (\ref{trpiF}) holds also true in the $T$ invariant
case, as is clear from Eqs. (\ref{gpi},\ref{gFkt}).
Eqs. (\ref{JPP},\ref{trpiF}) confirm our hypotheses
that the {\it parity} polarization current is indeed
robust against $I$ symmetry.

\sc{Quantum spin pump}

\ind
Another possible direction to be explored is to study how to induce 
a spin polarization current by breaking both $T$-invariance 
and $I$-invariance. This scenario can be implemented \cite{RS2} 
in a certain kind of quantum spin chains such as Cu-bensoate and 
${\rm Yb}_{4}{\rm As}_{3}$.  
The ground state of these quantum magnets 
is known to be quantum critical point (QCP), 
which is interpreted as a Dirac monopole, i.e., a source of 
the U(1) field strength  ${\cal F}_{kt}$. 
When this quantum system is driven {\it around} this 
QCP by applying an electric field ${\bf E}$ and/or 
magnetic field ${\bf B}$, a spin polarization current can 
be induced.   The electromagnetic fields break 
both the $I$-invariance and $T$-invariance. They also  
induce a spin gap and realize a quantum critical point 
at the origin of ${\bf E}$-${\bf B}$ plane. When the system  
goes adiabatically around this origin, a quantized number of 
spins will be transported from one edge to the other through the 
system. This quantized value is a physical 
manifestation of the first Chern number associated with the QCP.

\sb{${\cal F}_{k_\mu x_\mu}$ associated with the spatial inhomogeneity}

\ind
Contrary to ${\cal F}_{k_{\mu}k_{\nu}}$ and ${\cal F}_{k_{\mu}t}$,
the reciprocal field strength ${\cal F}_{k_\mu x_\nu}$ 
does not seem to be related directly to a physical observable 
such as Hall type currents and polarization currents. 
However, when the system contains spatial inhomogeneity such as 
lattice defects, ${\cal F}_{k_{\mu}x_{\nu}}$ appears  
and plays an important role in the dynamics of Bloch electrons 
around the defects \cite{SN}. Another possible application of 
${\cal F}_{k_\mu x_\nu}$ is the electron transport properties 
around a magnetic domain wall, where the spatial modulation of 
ferromagnetic moments induce 
${\cal F}_{k_{\mu}x_{\nu}}$,
and naturally influences the EOM for 
the electron wave packet through this Berry curvature.

\sa{Conclusions}

\ind
We have derived and analyzed the semiclassical EOM 
for a wave packet of Bloch electrons, 
under perturbations slowly varying in space and in time.
Their interpretation in terms of non-abelian gauge field
in the reciprocal parameter space was the central issue of
the paper.
The same type of EOM has been previously derived for the
abelian, i.e., U(1) case, by using either
(i) Time-dependent variational principle \cite{SN} or
(ii) Path-integral method using Wannier basis \cite{KT}.
We have generalized such EOM to a non-abelian case
by using only the most fundamental principles
of quantum mechanics.

The advantage of our formalism was that
\ben
\item
it was {\it asymptotically exact} in the framework of linear 
response theory, as a result of systematic expansion
w.r.t. the perturbation $\beta$ or ${\bf x}-\bar{\bf x}$,
\item
it revelaed that there are different types of gauge field 
of different physical origin,
\item
it was useful for developing symmetry analyses on various types
of Berry phase transport.
\een 
The first point refers to Eq. (\ref{H1}) and all the related
analyses developped in Secs. 3 and 4. 
The relevance of our results
in relation to other approaches was further discussed in Sec. 4.4.
As for the second point, two different sources of gauge
field have been revealed, i.e.,
(i) Projection onto a subspace spanned by $N$ Bloch bands,
(ii) Bloch basis moving in the course of time.
The former is the origin of ${\cal F}_{k_\mu k_\nu}$
which is directly related to spontaneous Hall currents
of various degrees of freedom. 
The latter brings about ${\cal F}_{k_\mu x_\mu}$ and 
${\cal F}_{k_\mu t}$, which plays an important role in the spatially
and temporally inhomogeneous system. 

Finally, concerning the last point in the above list,
we have applied our formalism to the analyses on the
spin and orbital transport phenomena with the help of
Boltzmann transport theory.
The role of time reversal and space inversion symmetries
in the appearence of finite Hall/polarization currents
has been extensively studied.
The {\it cancellation rules} are summarized in Tables 1 and 2.
The concept of {\it parity} polarization current has also
been introduced, which may concretize Berry phase engineering
in the context of {\it orbital} transport.

We leave for a future study further investigations on
their application to the domain wall physics and
that of quantum pumping. 
In conclusion, we believe that our analyses on non-Abelian gauge 
field will see in the near future a possible application
in the context of Berry phase engineering.

\vspace{0.2cm}
Note added - After completion of this work, we were informed
of a related effort by D. Culcer, Y. Yao and Q. Niu.
\cite{CYN}

\vspace{0.5cm}
\bc
{\large\bf Acknowledgements}
\ec

\vspace{0.2cm}
We would like to thank Shuichi Murakami and Naoto Nagaosa for introducing 
us into this flourishing area of physics, as well as giving us
a motivation to work on this problem.
We are also grateful to Dimitrie Culcer and Qian Niu for their comments
and suggestions on our paper.

R.S. is a JSPS Postdoctoral Fellow. K.I. is supported by RIKEN as 
a Special Postdoctoral Researcher.

\vspace{0.5cm}
\noindent
{\large\bf Appendix A : 
Matrix element $\bla H\bra_{mn}({\bf k})$
and $\Delta\Omega^{(1)}_{x_{\mu}}$}

\vspace{0.1cm}
Matrix elements of the linearlized Hamiltonian, i.e.,
Eqs. (\ref{Hmn1},\ref{Hmn2}),
have been extensively used in Sec. 4.3.
Let us recall those equations together
with the matrix elements of noncommutative coordinates,
i.e., Eqs. (\ref{xmn1},\ref{xmn2},\ref{xx},\ref{dd}).
Our purpose here is to substitute the expression (\ref{Hmn2}) 
into 
\be
\Omega_{x_\mu}^{(1)}=
i\sum_{m,n=1}^N
\int d{\bf k} a_{m}^{\ast}({\bf k})
\Big[\bla H\bra, \bla x_\mu\bra\Big]_{mn}({\bf k})
a_n({\bf k}),
\label{omgx1}
\ee
an expression analogous to the second line of Eqs. (\ref{dec}), 
and to rewrite $\Omega^{(1)}_{x_{\mu}}$ in terms of the field 
strength $\big[{\cal F}_{k_\mu k_\nu}\big]_{mn}$.

The first term of Eq. (\ref{Hmn2}), i.e.,
$\epsilon_{\rm loc}({\bf k},\bar{\bf x},t)$ gives a standard velocity
term when inserted into Eq. (\ref{omgx1}).
Since $\big[{\cal F}_{k_\mu k_\nu}\big]_{mn}$ is related
to the commutator, $\big[\nabla_{k_\mu},\nabla_{k_\nu} \big]_{mn}$
or equivalently, $\big[\bla x_\mu\bra, \bla x_\nu\bra\big]_{mn}({\bf k})$,
one can easily imagine that the last two terms give in
Eq. (\ref{Hmn2}) give when inserted into Eq. (\ref{omgx1})
a contribution related to $\big[{\cal F}_{k_\mu k_\nu}\big]_{mn}$.
One can indeed verify
\bea
&&-\Big[{1\over 2}\Big(
{\rd \epsilon_{\rm loc}\over\rd\bar{x}_\nu}\nabla_{k_{\nu}}+
\nabla_{k_{\nu}}{\rd \epsilon_{\rm loc}\over\rd\bar{x}_\nu}
\Big),i\nabla_{k_{\mu}}\Big]_{mn}
\nn \\
&=&-\frac{i}{2}\bigg\{
{\rd \epsilon_{\rm loc}\over\rd\bar{x}_\nu}
\Big[\nabla_{k_{\nu}},\nabla_{k_{\mu}}\Big]_{mn}
+\sum_{l=1}^{N}\Big[{\rd \epsilon_{\rm loc}\over\rd\bar{x}_\nu},
\nabla_{k_{\mu}}\Big]_{ml}\Big[\nabla_{k_{\nu}}\Big]_{ln}
\nn \\
&+&\sum_{l=1}^{N}\Big[\nabla_{k_{\nu}}\Big]_{ml}
\Big[{\rd \epsilon_{\rm loc}\over\rd\bar{x}_\nu},
\nabla_{k_{\mu}}\Big]_{ln}+
\Big[\nabla_{k_{\nu}},\nabla_{k_{\mu}}\Big]_{mn}
{\rd \epsilon_{\rm loc}\over\rd\bar{x}_\nu}
\bigg\}
\nn \\
&=&-{\rd \epsilon_{\rm loc}\over\rd\bar{x}_\nu}{\cal F}_{k_{\nu}k_{\mu}}+ 
\frac{i}{2}\bigg(\frac{\rd^{2} \epsilon_{\rm loc}}{\rd k_{\mu}\rd\bar{x}_\nu}
\Big[\nabla_{k_{\nu}}\Big]_{mn}+
\Big[\nabla_{k_{\nu}}\Big]_{mn}
\frac{\rd^{2} \epsilon_{\rm loc}}{\rd k_{\mu}\rd\bar{x}_\nu}\bigg).
\label{34}
\eea
The second term of Eq. (\ref{Hmn2}) gives, when inserted into
the commutator,
\be
\Big[\frac{\rd \epsilon_{\rm loc}({\bf k},\bar{\bf x},t)}
{\rd \bar{x}_{\nu}}\bar{x}_{\nu},
\nabla_{k_{\mu}}\Big]_{mn}=-
\frac{\rd^{2} \epsilon_{\rm loc}}{\rd k_{\mu}
\rd \bar{x}_{\nu}}\bar{x}_{\nu}.
\label{2}
\ee
Collecting the contribution (\ref{2}) and the last two terms of 
Eq. (\ref{34}), i.e., terms not related to ${\cal F}_{k_{\nu}k_{\mu}}$,
one defines $\Delta \Omega^{(1)}_{x_{\mu}}$ introduced
in Eq. (\ref{eff}) ;
\be
\Delta \Omega^{(1)}_{x_{\mu}}=
-\sum_{\nu,m,n}\int d{\bf k}
\frac{\rd^{2}\epsilon_{\rm loc}}{\rd k_{\mu} \rd \bar{x}_{\nu}}
\rho({\bf k},t)\bigg\{
\bar{x}_{\nu}(t)-iz^{\ast}_{m}({\bf k},t)\Big[\nabla_{k_{\nu}}\Big]
z_{n}({\bf k},t)
\bigg\}.
\label{irr}
\ee
This term vanishes after $k$-integration with the help of
prescription given in Eq. (\ref{presc}).

\vspace{0.5cm}
\noindent
{\large\bf Appendix B : Derivation of the EOM for $\bar{\bf z}(t)$}

\vspace{0.1cm}
We demonstrate here the derivation of EOM for $\bar{\bf z}(t)$,
i.e., EOM describing the motion of the internal pseudospin degree 
of freedom.
For that purpose we once have to go back to Eq. (\ref{dadt}). 
After multiplying it with a weight $\sqrt{\rho({\bf k},t)}$,
we integrate it over all the ${\bf k}$-points, to find,
\bea
&&
\int d{\bf k}
\bigg\{
\frac{1}{2}\rdt\big(\rho({\bf k},t)\big){\bf z}({\bf k},t)+\rho
\frac{\rd {\bf z}}{\rd t} 
\bigg\} 
\nn \\
&=&i\int d{\bf k}\rho({\bf k},t)
\bigg\{-\epsilon_{\rm eff}+
\bar{x}_{\mu}\frac{\rd \epsilon_{\rm loc}}{\rd \bar{x}_{\mu}}-
\frac{\rd\epsilon_{\rm loc}}{\rd \bar{x}_{\mu}}
A_{k_{\mu}}+
\frac{d\bar{x}_{\mu}}{dt}{\cal A}_{\bar{x}_{\mu}}+ 
{\cal A}_{t}
\bigg\}{\bf z} 
\nn \\
&+&\frac{1}{2}\int d{\bf k}\rho({\bf k},t)
\frac{\rd \epsilon_{\rm loc}}{\rd \bar{x}_{\mu}}
\frac{\rd {\bf z}}{\rd k_{\mu}},
\label{dzdt1}
\eea
where 
$\epsilon_{\rm eff}=
\epsilon_{\rm loc}{\bf 1}+\Delta \epsilon$
as given in Eq. (\ref{eff}).
In Eq. (\ref{dzdt1}),
repeated $\mu$-indices should be summed over
$\mu=1,\cdots,D$.
In order to obtain Eq. (\ref{dzdt1}), we also used,
\be
\int d{\bf k} \bigg\{\sqrt{\rho}
\frac{\rd \epsilon_{\rm loc}(\bar{\bf k})}
{\rd \bar{x}_{\mu}}\frac{\rd}{\rd k_{\mu}}
\Big(\sqrt{\rho}{\bf z}\Big) + \frac{1}{2}\rho{\bf z}
\frac{\rd \epsilon_{\rm loc}}{\rd k_{\mu} \rd \bar{x}_{\mu}} \bigg\}
=\frac{1}{2}\int d{\bf k} \rho
\frac{\rd \epsilon_{\rm loc}}{\rd \bar{x}_{\mu}}
\frac{\rd {\bf z}}{\rd k_{\mu}}
\label{dzdq}.
\ee
We now substitute,
\[
\rdt\big(\rho({\bf k},t)\big)
=\sum_{\mu=1}^{D}\frac{\rd}{\rd k_{\mu}}\bigg\{
\rho({\bf k},t)
\frac{\rd \epsilon_{\rm loc}({\bf k},\bar{\bf x},t)}
{\rd \bar{x}_{\mu}}\bigg\},
\]
into Eq. (\ref{dzdt1}), then perform a partial integral 
w.r.t. $k_{\mu}$. The result is,
\bea
&&\int d{\bf k}\rho ({\bf k},t)\bigg\{-
\frac{\rd \epsilon_{\rm loc}}{\rd \bar{x}_{\mu}}
\frac{\rd {\bf z}}{\rd k_{\mu}}+ 
\frac{\rd {\bf z}}{\rd t}\bigg\}
\nn \\
&&=i\int d{\bf k}\rho\bigg\{- 
\epsilon_{\rm eff}+
\bar{x}_{\mu}\frac{\rd \epsilon_{\rm loc}}{\rd \bar{x}_{\mu}}-
\frac{\rd\epsilon_{\rm loc}}{\rd \bar{x}_{\mu}}
A_{k_{\mu}}+
\frac{d\bar{x}_{\mu}}{dt}{\cal A}_{\bar{x}_{\mu}}+ 
{\cal A}_{t}
\bigg\}{\bf z}.
\label{dzdt2}
\eea
Finally,  
in order to rewrite Eq. (\ref{dzdt2}) in the form of Eq. (\ref{EOM2})
and complete its derivation,
we adopt the prescritpion (\ref{presc}).

\vspace{0.5cm}
\noindent
{\large\bf Appendix C : EOM for 
${\cal O}_{\mu\alpha}^{\cal S}$ and spin Hall current}

\vspace{0.1cm}
Our purpose here is to rewrite an EOM for
Eq. (\ref{SH1}) into its final form,
i.e., Eq. (\ref{SH2}), so that we can express
the spin Hall current as the following trace 
in the pseudospin space,
${\rm Tr}\big[\la S_\alpha\ra(\bar{\bf k})
{\cal F}_{\bar{k}_\mu \bar{k}_\nu}(\bar{\bf k})\big] E_{\nu}$.

Let us first recall the assumption we made in Eq. (\ref{as1}). 
This assumption allows us to factorize
$\bla {\cal O}_{\mu\alpha}^{\cal S}\bra_{mn}
({\bf k},{\bf k}^{\prime})$ 
in Eq. (\ref{SH1}) into a product of 
$\la S_\alpha\ra$ and  $\la x_\mu\ra$;
\[
\bla S_\alpha x_\mu +x_\mu S_\alpha \bra_{mn}= 
\sum_{l=1}^N\Big(
\la S_\alpha\ra_{ml}
\la x_\mu\ra_{ln}+\la x_\mu\ra_{ml} 
\la S_\alpha\ra_{ln}\Big).
\] 
Correspondingly, the commutator appearing in Eq. (\ref{SH1}) 
can be decoupled into the following two types of commutators,  
\bea
\Big[\l\la H \r\ra,\l\la {\cal O}_{\mu\alpha}^{\cal S}\r\ra \Big]
=\frac{1}{2}
\Big[\bla H\bra,\bla S_\alpha\bra\Big]\bla x_{\mu}\bra +
\frac{1}{2}
\bla S_\alpha\bra \Big[\bla H \bra ,\bla x_{\mu}\bra \Big]
-{\rm h.c.}
\label{dec2}
\eea
The first term together with its hermitian conjugate 
conceives a commutator between $\l\la H\r\ra$  and $\l\la S_\alpha\r\ra$, 
which consitutes the EOM for the spin : $\frac{dS_\alpha}{dt}$. 
Firstly, we show that this commutator vanishes.
Since $\l\la S_\alpha\r\ra $ is diagonal in ${\bf k}$-space, 
it clearly commutes with $\l\la H_{0}\r\ra$;
\bea
\Big[\l\la H_{0}\r\ra, \l\la S_\alpha\r\ra\Big]_{mn}
({\bf k}^{\prime},{\bf k})&=&\delta({\bf k}^{\prime}-{\bf k}) 
\Big[\epsilon_{\rm loc}({\bf k}),\l\la S_\alpha\r\ra ({\bf k})\Big]_{mn}
\nn \\
&=&0.
\eea 
Therefore, the commutator between $\l\la H\r\ra$ 
and $\l\la S_\alpha\r\ra$ becomes proportional to 
the covariant derivative of $\l\la S_\alpha\r\ra({\bf k})$ 
w.r.t. $k_{\mu}$, i.e.,
\bea
&&\Big[\l\la H\r\ra ,\l\la S_\alpha\r\ra\Big]_{mn}({\bf k}^{\prime},{\bf k})
=e \sum_{\mu=1}^{D} E_{\mu}
\Big[\l\la x_{\mu}\r\ra  ,\l\la S_\alpha\r\ra\Big]_{mn}
({\bf k}^{\prime},{\bf k})\nn \\
&&=ie\ \delta({\bf k}^{\prime}-{\bf k})\sum_{\mu=1}^{D}
E_{\mu} \Big[\nabla_{k_{\mu}},
\l\la S_\alpha\r\ra({\bf k})\Big]_{mn}.
\label{dS1}
\eea
Because the spin operator 
itself does not depend on the crystal momentum, 
the partial derivative of $\l\la S_\alpha\r\ra({\bf k})$ 
w.r.t. $k_{\mu}$ reduces to 
the commutator between $i {\cal A}_{k_{\mu}}$ 
and $\l\la S_\alpha\r\ra({\bf k})$, : 
\bea
&&{\rd\over\rd k_{\mu}}\bla S_\alpha\bra_{mn} ({\bf k})=
\Bla\frac{\rd u_{m}({\bf k})}{\rd k_{\mu}}
\Big|S_\alpha\Big|u_{n}({\bf k})\Bra 
+ \Bla u_{m}({\bf k})\Big|S_\alpha
\Big|\frac{\rd u_{n}({\bf k})}{\rd k_{\mu}}\Bra
\nn \\ 
&&=
\sum_{l=1}^N\Big\{
\Bla\frac{\rd u_{m}}{\rd k_{\mu}}\Big|u_{l}\Bra 
\Bla u_{l}\Big|S_\alpha\Big|u_{n}\Bra
+\Bla u_{m}\Big|S_\alpha\Big|u_{l}\Bra 
\Bla u_{l} \Big|\frac{\rd u_{n}}{\rd k_{\mu}}\Bra\Big\}.
\label{DtS}
\eea 
In the second line $k$-dependence is not written explicitly. 
We used Eq. (\ref{as1}) between the two lines. 
Then the covariant derivative of $\l\la S_\alpha\r\ra ({\bf k})$ 
w.r.t. $k_{\mu}$ appearing in Eq. (\ref{dS1}) also vanishes,   
\bea
\Big[\nabla_{k_{\mu}},\l\la S_\alpha\r\ra({\bf k})\Big]_{mn} 
&=&\frac{\rd}{\rd k_{\mu}}\Big(\l\la S_\alpha\r\ra_{mn}({\bf k})\Big)- 
\Big[i{\cal A}_{k_{\mu}},\l\la S_\alpha\r\ra({\bf k})\Big]_{mn} 
\nn \\
&=&0. 
\label{dS2}
\eea
Consequently, the first term and its hermitian conjugate in 
Eq. (\ref{dec2}) are indeed zero. On the other hand, 
the second term in  Eq. (\ref{dec2}) contains the commutator 
between $\l\la H\r\ra$ and $\l\la x_{\nu}\r\ra$, which 
describes the EOM for $x_{\nu}$ now.  This term gives rise to 
a field strength ${\cal F}_{k_{\mu}k_{\nu}}$ through 
the commutator between covariant derivatives 
w.r.t. different components of the crystal momentum, 
i.e., $\big[\nabla_{k_{\mu}},\nabla_{k_{\nu}} \big]$.
Namely,
\bea
\Big[\l\la H\r\ra, \l\la x_{\nu}\r\ra\Big]
&=&\Big[\l\la H_{0}\r\ra + e\ \sum_{\mu=1}^{D}E_{\mu}  
\l\la x_{\mu}\r\ra,
\l\la x_{\nu}\r\ra\Big]({\bf k}^{\prime},{\bf k})\nn \\ 
&=&\delta({\bf k}^{\prime}-{\bf k})\ 
\Big[\epsilon_{\rm loc}({\bf k}){\bf 1} + e\sum_{\mu=1}^{D}
E_{\mu}\ i\nabla_{k_{\mu}},
i\nabla_{k_{\nu}}\Big] \nn \\
&=&-i\delta({\bf k}^{\prime}-{\bf k}) 
\bigg\{
{\rd \epsilon_{\rm loc}\over\rd k_\nu}{\bf 1}-
e\sum_{\mu=1}^{D}E_{\mu}
\Big({\cal F}_{k_{\mu}k_{\nu}}
({\bf k})\Big)
\bigg\}.
\label{Dx}
\eea
Finally, substituting Eq. (\ref{dec2}) 
together with Eqs. (\ref{dS1},\ref{dS2}) and (\ref{Dx})
into Eq. (\ref{SH1}), one finds  Eq. (\ref{SH2}).

\vspace{0.5cm}
\noindent
{\large\bf Appendix D : EOM for ${\cal O}_\mu^\Pi$ and 
parity polarization current}

\vspace{0.1cm}
In parallel with Appendix C, we rewrite below Eq. (\ref{PP1})
into its final form, i.e., Eq. (\ref{PP2}).
Let first recall the assumption (\ref{as2}), which says that
the parity operator $\Pi$ has no matrix element 
outside the $N$-fold degenerate band.
This implies that
$\l\la {\cal O}_\mu^\Pi  \r\ra$ can be factorized into 
a product of two $N$ by $N$ matrices, or
$\la\Pi x_\mu\ra=\la\Pi\ra \la x_\mu\ra$.
Thanks to this factorization,
the commutator in Eq. (\ref{PP1}) can be 
decomposed into two types of commutators as
\bea
\Big[\bla H_{\rm loc}\bra-i\nabla_{t}, 
\bla {\cal O}_\mu^\Pi\bra\Big]
&=&
\frac{1}{2}\Big(
\Big[\bla H_{\rm loc}\bra-i\nabla_{t}, 
\bla\Pi\bra \Big]\bla x_{\mu}\bra
+{\rm h.c.}\Big)  
\nn \\
&+&\frac{1}{2}\Big(
\la\Pi\ra\Big[\l\la H_{\rm loc}\r\ra-i\nabla_{t},
\bla x_\mu\bra\Big]+{\rm h.c.}\Big)  
\label{D1}
\eea
On the r.h.s., the first term is a commutator 
between $\l\la H_{\rm loc}\r\ra  -i \nabla_{t}$ and $\l\la \Pi \r\ra$, 
which constitutes the EOM of the parity under time-dependent 
perturbations: $d\Pi/dt$.  Since the parity operator 
itself is independent of time, we can prove that 
this commutator indeed vanishes, 
in the same way as we did in Eq. (\ref{DtS}),  
\bea
\big[\l\la H_{\rm loc}\r\ra -i \nabla_{t}, \l\la \Pi \r\ra \big] = 0.
\label{D2} 
\eea   
On the other hand, the second line of Eq. (\ref{D1}) 
is a commutator between 
$\la H_0\ra -i \nabla_{t}$ and $\la x_\mu\ra$, 
which describes the EOM for $x_\mu$. 
This commutator gives rise to ${\cal F}_{k_{\nu}t}$ 
through a commutation relation  
between two covariant derivatives, 
one w.r.t. time and the other w.r.t. the momentum,
$\big[i\nabla_{k_{\nu}},i\nabla_{t}\big]
= i{\cal F}_{k_{\nu}t}$.
Thus the second line of Eq. (\ref{D1}) may be rewritten
as
\bea
&&\Big[\la H_{\rm loc}\ra-i\nabla_{t},
\la x_\mu\ra\Big]_{mn}
({\bf k}^{\prime},{\bf k}) 
=i\delta({\bf k}^{\prime}-{\bf k})
\Big[\epsilon_{\rm loc}({\bf k}) - i\nabla_{t},i\nabla_{k_\mu}
\Big]_{mn} \nn \\ 
&=& -i\delta({\bf k}^{\prime}-{\bf k})\  
\bigg\{\frac{\rd \epsilon_{\rm loc}({\bf k})}{\rd k_{\nu}} \delta_{mn} 
- \Big({\cal F}_{k_\mu t}\Big)_{mn}\bigg\}. 
\label{D3}
\eea
Substituting Eq. (\ref{D1}) together with Eqs. 
(\ref{D2},\ref{D3}) into Eq. (\ref{PP1}), 
one finds Eq. (\ref{PP2}).

\vspace{1.0cm}
\bb

\bit{dirac}
P.A.M. Dirac, Proc. R. Soc. London 133, 60 (1931).

\bit{pol}
G. 't Hooft, Nucl. Phys. B 79, 276 (1974);
A.M. Polyakov, JETP Lett. 20, 194 (1974).

\bit{masa}
M. Onoda, N. Nagaosa, J. Phys. Soc. Jpn. 71, 19 (2002).

\bit{fang}
Z. Fang, et al., Science 302, 92 (2003).

\bit{vol}
See also, G.E. Volovik, JETP Lett. 46, 98 (1987),
and references therein.

\bit{CN}
M.C. Chang, Q. Niu, Phys. Rev. Lett. 75, 1348 (1995);
ibid., Phys. Rev. B 53, 7010 (1996).

\bit{SN}
G. Sundaram, Q. Niu, Phys. Rev. B 59, 14915 (1999).

\bit{hirsch}
J.E. Hirsch, Phys. Rev. Lett. 83, 1834 (1999).

\bit{zhang}
S. Zhang, Phys. Rev. Lett. 85, 393 (2000).

\bit{MNZ}
S. Murakami, N. Nagaosa, S.C. Zhang, Science 301, 1348 (2003);
ibid., Phys. Rev. B 69, 235206 (2004).

\bit{sinova}
J. Sinova, et al., Phys. Rev. Lett. 92, 126603 (2004).

\bit{culcer}
D. Culcer, et al., Phys. Rev. Lett. 93, 046602 (2004).

\bit{DD}
S. Datta, B. Das, Appl. Phys. Lett. 56, 665 (1990).

\bit{TN}
D. J. Thouless, Phys. Rev. B 27, 6083 (1983);
Q. Niu, Phys. Rev. Lett. 64, 1812 (1990).

\bit{RS2}
R. Shindou, cond-mat/0312668.

\bit{berry}
M.V. Berry, Proc. R. Soc. London, A 392, 45 (1984).

\bit{mead}
C.A. Mead, Chem. Phys. 49, 23 (1980)

\bit{KL}
R. Karplus, J.M. Luttinger, Phys. Rev. 95, 1154 (1954).

\bit{zak}
J. Zak, Phys. Rev. B 15, 771 (1977); ibid., 16, 4154 (1977).

\bit{blout}
E.N. Adams and E.I. Blout, Phys. and Chem. Solids 10, 286, (1959);
E.I. Blout, Solid State Phys. 13, 305, (1962).

\bit{TKNN}
D.J. Thouless, et al.,
Phys. Rev. Lett. 49, 405 (1982).

\bit{resta}
R.D. King-Smith, D. Vanderbilt,
Phys. Rev. B 47, 1651 (1993);
R. Resta, Europhys. Lett. 22, 133 (1993).
See also,
R. Resta, Rev. Mod. Phys. 66, 899 (1994),
and references therein.

\bit{OMN}
K. Ohgushi, S. Murakami, N. Nagaosa,
Phys. Rev. B 62, 6065 (2000).

\bit{RS1}
R. Shindou, N. Nagaosa, Phys. Rev. Lett. 87, 116801 (2001).

\bit{JNM}
T. Jungwirth, Q. Niu, A.H. MacDonald,
Phys. Rev. Lett. 88, 207208 (2002).

\bit{yao}
Y. Yao et al. Phys. Rev. Lett. 92, 037204 (2004).

\bit{WZ}
F. Wilczek, A. Zee Phys. Rev. Lett. {\bf 52}, 2111 (1984).

\bit{inoue}
J. Inoue, G.E.W. Bauer, L.W. Molenkamp, Phys. Rev. B 70,
R041303 (2004).

\bit{loss}
J. Schliemann, D. Loss, Phys. Rev. B 69, 165315 (2004).

\bit{KT}
H. Koizumi, Y. Takada, Phys. Rev. B 65, 153104 (2002).

\bit{Metz}
A. Berard, H. Mohrbach, Phys. Rev. D 69, 127701 (2004).

\bit{seiberg}
N. Seiberg, E. Witten, J. High Energy Phys. 9909, 32 (1999).

\bit{connes}
A. Connes, M.R. Douglas, A. Schwarz,
J. High Energy Phys. 9802, 3 (1998).

\bit{vdB1}
N. Marzari, D. Vanderbilt, Phys. Rev. B 56, 12847 (1997).

\bit{kenzo}
K. Ishikawa, T. Matsuyama, Z. Phys. C 33, 41 (1986);
ibid., Nucl. Phys. B 280, 523 (1987).

\bit{SCZ}
S.C. Zhang, T.H. Hansson, S. Kivelson, Phys. Rev. Lett. 
62, 82 (1989). See also, S.C. Zhang, Int. J. Mod. Phys.
6, 25 (1992).

\bit{goryo}
G.E. Volovik, JETP 67, 1804 (1988);
J. Goryo, K. Ishikawa, Phys. Lett A 260, 294 (1999);
A. Furusaki, M. Matsumoto, M. Sigrist, Phys. Rev. B 64, 
054514 (2001).

\bit{senthil}
G.E. Volovik, V.M. Yakovenko, J. Phys.: Condens. Matter 1, 
5263 (1989);
T. Senthil, J.B. Marston, M.P.A. Fisher, Phys. Rev. B 60, 
4245 (1999).

\bit{vdB2}
N. Sai, K.M. Rabe, D. Vanderbilt, Phys. Rev. B 66, 104108 (2002).

\bit{CYN}
D. Culcer, Y. Yao, Q. Niu, cond-mat/0411285.

\eb

\vfil\eject

\begin{table}[htbp]
\begin{center}
\begin{tabular}{|c|c|c|}
\hline
invariance & $\hspace{0.3cm}$ $T$ $\hspace{0.3cm}$ & 
             $\hspace{0.3cm}$ $I$ $\hspace{0.3cm}$ \\ 
\hline \hline
charge : ${\bf 1}$ & + & + \\ \hline 
spin : $\l\la S\r\ra({\bf k})$ & $-$ & + \\ \hline 
 parity : $\la \Pi\ra({\bf k})$ & + & $-$ \\ \hline \hline 
Hall-type current: ${\cal F}_{k_{\mu}k_{\nu}}({\bf k})$ 
& $-$ 
& $+$   \\  \hline
polarization current: ${\cal F}_{k_{\mu}t}({\bf k})$ & 
$+$  & 
$-$  \\ \hline 
\end{tabular}
\caption{Transformation properties under time reversal $T$ and 
spatial inversion $I$ -
A negative (positive) sign indicates whether 
a matrix element in question 
at ${\bf k}$ reverses its sign  
(or not) in comparison  with  that of $-{\bf k}$ when
the system is invariant under a certain symmetry operation,
such as $T$ or $I$. 
For example, the {\it negative} sign for $\la S\ra({\bf k})$
in the $T$-invariant case means that the spin matrix 
$\la S_\alpha\ra ({\bf k})$ is identical to  
$-\big(\l\la S_\alpha\r\ra (-{\bf k})\big)^{t}$ upto a certain 
$SU(N)$ gauge transformation as given in Eq.(\ref{gS}).}
\end{center}
\end{table}

\begin{table}[htbp]
\begin{center}
\begin{tabular}{|c|c|c|c|c|}
\hline 
type of current
& \multicolumn{2}{|c|}{Hall-type} 
& \multicolumn{2}{|c|}{polarization}\\ 
\hline
invariance & 
$\hspace{0.3cm}$ $T$ $\hspace{0.3cm}$ & 
$\hspace{0.3cm}$ $I$ $\hspace{0.3cm}$ & 
$\hspace{0.3cm}$ $T$ $\hspace{0.3cm}$ & 
$\hspace{0.3cm}$ $I$ $\hspace{0.3cm}$ \\
\hline \hline 
charge & $-$ & $+$ & $+$ & $-$  \\ 
\hline
spin & $+$ & $+$ & $-$ & $-$  \\ 
\hline
parity & $-$ & $-$ & $+$ & $+$  \\ \hline 
\end{tabular}
\caption{Cancellation rules for charge/spin/parity 
Hall-type/polarization currents -
A negative (positive) sign indicates that 
contributions to the total 
charge/spin/parity (vertical axis)
Hall-type/polarization (horizontal axis)
current from ${\bf k}$
and $-{\bf k}$ electrons (do not) cancel each other
when the system is invariant under either $T$ or $I$.
This table can be deduced from Table 1.
For example, in the $T$-invariant case, 
a {\it negative} sign for the spin in Table 1 gives, 
together with another {\it negative} sign for the Hall-type current,
a {\it positive} sign for the spin Hall current in Table 2,
correspoinding, respectively, to Eqs. 
(\ref{gS},\ref{gFkk}) and (\ref{trSF}).}  
\end{center}
\end{table}

\end{document}